\documentclass[aps,prx,twocolumn,amsmath,amssymb,superscriptaddress,longbibliography]{revtex4-2}

\usepackage{graphicx}
\usepackage{dcolumn}
\usepackage{bm}
\usepackage{amsmath,amssymb,amsfonts}
\usepackage{mathrsfs}
\usepackage{multirow}
\usepackage{booktabs}
\usepackage{xcolor}
\usepackage{textcomp}
\usepackage{empheq}

\usepackage[utf8]{inputenc}
\DeclareUnicodeCharacter{2212}{-}
\usepackage[T1]{fontenc}
\usepackage{mathptmx}
\usepackage{etoolbox}
\usepackage{hyperref}

\usepackage[english]{babel}
\renewcommand{\selectlanguage}[1]{}

\providecommand{\pab}[1]{\left( #1 \right)}

\begin{document}

\title{A near-quantum-limited diamond maser amplifier operating at millikelvin temperatures}

\author{Morihiro Ohta}
\thanks{These authors contributed equally to this work.}
\affiliation{Experimental Quantum Information Physics Unit, Okinawa Institute of Science and Technology Graduate University, Onna, Okinawa 904-0495, Japan}

\author{Ching-Ping Lee}
\thanks{These authors contributed equally to this work.}
\thanks{Present address: Department of Physics, National Tsing Hua University, Hsinchu 30013, Taiwan}
\affiliation{The Science and Technology Group, Okinawa Institute of Science and Technology Graduate University, Onna, Okinawa 904-0495, Japan}

\author{Vincent P.M. Sietses}
\thanks{Present address: QuTech and Kavli Institute of Nanoscience, Delft University of Technology, Delft, 2600 GA, The Netherlands}
\affiliation{Experimental Quantum Information Physics Unit, Okinawa Institute of Science and Technology Graduate University, Onna, Okinawa 904-0495, Japan}

\author{I. Kostylev}
\affiliation{The Science and Technology Group, Okinawa Institute of Science and Technology Graduate University, Onna, Okinawa 904-0495, Japan}

\author{J.R. Ball}
\thanks{Present address: Liquid Instruments, San Diego, CA 92130, USA}
\affiliation{Quantum Dynamics Unit, Okinawa Institute of Science and Technology Graduate University, Onna, Okinawa 904-0495, Japan}

\author{P. Moroshkin}
\thanks{Present address: School of Engineering, Brown University, Providence, RI 02912, USA}
\affiliation{Quantum Dynamics Unit, Okinawa Institute of Science and Technology Graduate University, Onna, Okinawa 904-0495, Japan}

\author{T. Hamamoto}
\affiliation{Experimental Quantum Information Physics Unit, Okinawa Institute of Science and Technology Graduate University, Onna, Okinawa 904-0495, Japan}

\author{Y. Kobayashi}
\affiliation{Sumitomo Electric Industries Ltd., Itami, Hyogo 664-0016, Japan}

\author{S. Onoda}
\affiliation{Quantum Materials and Applications Research Center, Takasaki Institute for Advanced Quantum Science, National Institutes for Quantum Science and Technology, Takasaki, Gunma 370-1292, Japan}

\author{T. Ohshima}
\affiliation{Quantum Materials and Applications Research Center, Takasaki Institute for Advanced Quantum Science, National Institutes for Quantum Science and Technology, Takasaki, Gunma 370-1292, Japan}
\affiliation{Department of Materials Science, Tohoku University, Sendai, Miyagi 980-8579, Japan}

\author{J. Isoya}
\affiliation{Graduate School of Pure and Applied Sciences, University of Tsukuba, Tsukuba, Ibaraki 305-8550, Japan}

\author{Hiroki Takahashi}
\affiliation{Experimental Quantum Information Physics Unit, Okinawa Institute of Science and Technology Graduate University, Onna, Okinawa 904-0495, Japan}

\author{Yuimaru Kubo}
\email[To whom correspondence should be addressed;\\ E-mail: ]{yuimaru.kubo@oist.jp}
\affiliation{The Science and Technology Group, Okinawa Institute of Science and Technology Graduate University, Onna, Okinawa 904-0495, Japan}

\date{\today}

\begin{abstract}
Current microwave quantum technologies require the amplification of weak signals with
minimal added noise at millikelvin temperatures. 
To date, this stringent requirement has been met exclusively by superconducting technologies, such as Josephson or kinetic-inductance parametric amplifiers. 
A fundamentally distinct alternative approach could be offered by masers, the microwave counterpart of lasers, which were predicted as early as the 1950s to achieve quantum-limited noise performance under ideal conditions. 
However, their dependence on cryogenic operation historically limited further advancement. 
Here we demonstrate the first-ever non-superconducting, near-quantum-limited maser amplifier operating at millikelvin temperatures utilising nitrogen impurity spins (P1 centres) in diamond. 
Population inversion is achieved via microwave pumping, exploiting a four-spin cross-relaxation mechanism. 
We realise a maximum power gain exceeding $\mathrm{30\,dB}$, an added noise of approximately $2.55$ quanta above the standard quantum limit, and a maximum $\mathrm{1\,\text{dB}}$ output compression point of $\mathrm{-63\,dBm}$ at $\mathrm{6.595\,\text{GHz}}$. 
The ability to operate in strong static magnetic fields of arbitrary orientation may offer a complementary, non-superconducting route for applications such as semiconducting spin-qubit readout, magnetic-resonance spectroscopy, and dark-matter axion searches.
\end{abstract}

\maketitle


Measuring weak microwave signals from quantum devices under test at millikelvin temperatures is essential for a wide range of quantum applications~\cite{clerk_introduction_2010}. 
Over the past two decades, this challenge has been addressed through the tremendous advances in superconducting Josephson-junction-based resonant parametric amplifiers (RPAs) operating near~\cite{castellanos-beltran_widely_2007,yamamoto_fluxdriven_2008,hatridge_dispersive_2011,mutus_design_2013} 
or at the quantum limit of added noise~\cite{planat_understanding_2019,winkel_nondegenerate_2020}. 
In parallel, travelling-wave parametric amplifiers (TWPAs)~\cite{macklin_quantum_2015,white_traveling_2015,planat_photoniccrystal_2020},
which offer bandwidths of several gigahertz and high saturation power, have also witnessed rapid progress.
Recent TWPAs have achieved near-ideal quantum efficiency
\cite{chang_josephson_2025,wang_highefficiency_2025}. 
These amplifiers have enabled not only high-fidelity readout of superconducting qubits, but also readout of semiconducting qubits~\cite{stehlik_fast_2015,zheng_rapid_2019,schaal_fast_2020}, detection of weak spin echoes in magnetic resonance~\cite{bienfait_reaching_2016,bienfait_magnetic_2017,eichler_electron_2017}, and the measurement of mechanical motion in quantum optomechanics~\cite{clark_observation_2016,satzinger_quantum_2018,noguchi_qubitassisted_2017}. 
They are also increasingly employed in particle physics experiments, such as dark matter axion searches~\cite{brubaker_first_2017,backes_quantum_2021,admxcollaboration_search_2021}. 
Moreover, kinetic-inductance-based parametric amplifiers, both resonant and travelling-wave types, have also been remarkably developed over the past decade~\cite{eom_wideband_2012}, approaching the quantum limit~\cite{malnou_threewave_2021,parker_degenerate_2022} and demonstrating strong microwave squeezing 
under magnetic fields approaching $2\,\mathrm{T}$~\cite{vaartjes_strong_2024}.

Whilst these superconducting devices have so far remained the leading--and indeed the only--practical solution for such demanding applications, masers (microwave amplification by stimulated emission of radiation) have been recognised as another fundamental method for microwave amplification since the 1950s. 
Theoretical work at that time already established that masers could reach quantum-limited noise under ideal conditions--that is, in the absence of thermal noise and with sufficient population inversion~\cite{weber_maser_1957,shimoda_fluctuations_1957,siegman_microwave_1964}. 
Despite this early promise, however, solid-state masers have been constrained by their dependence on cryogenic operation, which has hindered their application in classical communication technologies. 
Yet in the context of modern microwave quantum technologies, the cryogenic requirements are no longer a drawback. 

Optically-pumped diamond masers based on negatively charged nitrogen-vacancy ($\text{NV}^{-}$) centres have emerged as both oscillators~\cite{jin_proposal_2015,breeze_continuouswave_2018} and amplifiers at room temperature~\cite{day_roomtemperature_2024} and cryogenic temperatures~\cite{sherman_diamondbased_2022}. 
In addition, cavity cooling~\cite{ng_quasicontinuous_2021,fahey_steadystate_2023,day_roomtemperature_2024} and enhanced magnetometry~\cite{eisenach_cavityenhanced_2021} at room temperature have been demonstrated, suggesting the potential of diamond-based masers as resources for microwave quantum technology applications. 
Although optically pumped NV-based masers could in principle operate at millikelvin temperatures with a few microwatts of pump power, optical pumping remains challenging for integration with microwave quantum devices and requires precise alignment of the magnetic field along one of the $\langle111\rangle$ NV axes. 

Here we achieve population inversion in substitutional nitrogen defects (P1 centres) in diamond through microwave pumping at power levels of $100$ nanowatts or less, and demonstrated amplification with near-quantum-limited noise at millikelvin temperatures. 
It should be noted that the lowest previously reported noise temperature of $1.43\,\mathrm{K}$ at $8.4\,\mathrm{GHz}$ was achieved in a ruby (Cr$^{3+}$:Al$_2$O$_3$) maser operated at $1.6\,\mathrm{K}$, using a pump power of approximately $200\,\mathrm{mW}$ and a $1.6/295\,\mathrm{K}$ two-point Y-factor technique~\cite{glass_x-band_1994}.
This also suggests that masers might eventually serve as alternatives to higher-temperature HEMT (high-electron-mobility transistor) amplifiers, provided substantially broader bandwidths can be achieved. 

\subsection*{Maser amplifier device and operating principle} 

\begin{figure*}
    \centering
    \includegraphics[width=0.4\linewidth]{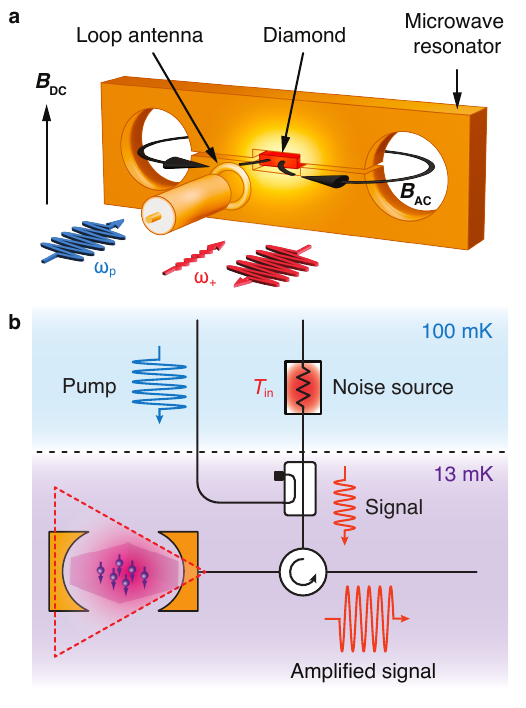}
    \caption{\textbf{Diamond maser amplifier device and setup.} 
    \textbf{a}, Device schematic. 
    A diamond crystal containing P1 centres is positioned within a loop-gap microwave resonator. 
    Both pump and probe microwave tones are introduced via a coupling loop antenna. 
    \textbf{b}, Experimental setup at the cold plate ($100\,\text{mK}$) and mixing chamber plate ($13\,\text{mK}$) in a dilution refrigerator.
    The resonator is enclosed in a copper housing and thermalised to the mixing chamber plate. 
    Amplified microwave signals are routed to the measurement line through a series of two circulators. 
    Additionally, a temperature-variable noise source, weakly thermalised to $100\,\text{mK}$, generates input noise characterised by a temperature $T_{\text{in}}$. 
    }
    \label{fig:Fig1}
\end{figure*}
\begin{figure*}
    \centering
    \includegraphics[width=\textwidth]{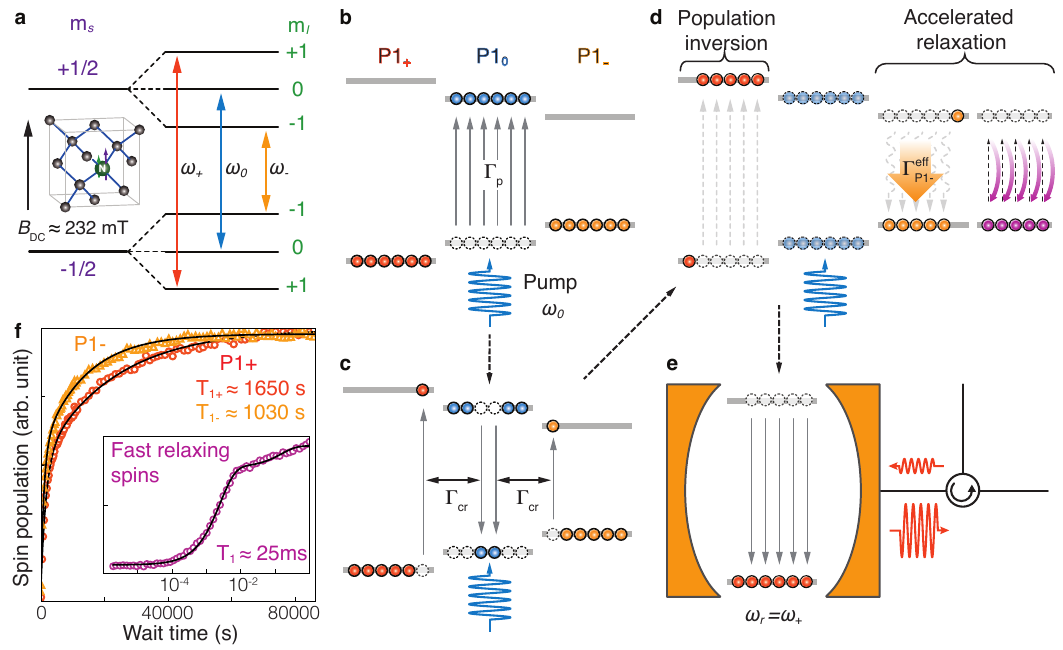}
    \caption{\textbf{Operating principle of the diamond maser amplifier.} 
    \textbf{a}, Illustration of a P1 centre in diamond and its energy levels under a static magnetic field of $B_\text{DC} \approx 232\,\text{mT}$, aligned parallel to the $[001]$ crystallographic axis. 
    \textbf{b},\textbf{c},\textbf{d},\textbf{e}, System dynamics with a pump. 
    \textbf{b}, A microwave pump tone at $\omega_0$ excites P1$_0$ spins at a rate of $\Gamma_\text{p}$. 
    \textbf{c}, This pump induces the excitation of one P1$_{+}$ and one P1$_{-}$ spin by exchanging energy with two P1$_0$ spins at a rate $\Gamma_{\text{cr}}$ via four-spin cross-relaxation. 
    \textbf{d}, Most P1$_{-}$ spins quickly deplete to the lower energy state due to the ``fast-relaxing spins'' (detailed further in \textbf{f}). 
    Consequently, only P1$_{+}$ spins progressively accumulate in the higher energy state, culminating in population inversion in the P1$_+$ transition. 
    \textbf{e}, The inverted P1$_{+}$ state amplifies the input microwave probe tone through stimulated emission. 
    \textbf{f}, Asymmetry in energy relaxation times between P1$_{+}$ and P1$_{-}$, indicating a faster decay in P1$_{-}$. 
    The inset shows the energy relaxation of the ``fast-relaxing spins'' on a timescale of approximately ten milliseconds. 
    While the data in the main panel were extracted from reflection spectroscopy measurements, those in the inset were obtained using a saturation recovery pulse sequence (see Supplementary Information section 2.1).
    }
    \label{fig:Fig2}
\end{figure*}

Figure~\ref{fig:Fig1}a illustrates the maser amplifier device, 
consisting of a loop-gap microwave resonator~\cite{ball_loopgap_2018} that houses a diamond crystal containing approximately $16\,\text{ppm}$ of P1 centres
~\cite{grezes_multimode_2014}. 
The resonator's frequency is $\omega_r/2\pi \approx 6.595\,\text{GHz}$, with internal and external quality factors of $Q_\text{int} \approx 2600$ and $Q_\text{ext} \approx 250$ (a total quality factor of $Q_\text{tot} \approx 230$).
The device is thermalised to the mixing chamber plate at $13\,\text{mK}$ in a dilution refrigerator. 
As shown in Figure~\ref{fig:Fig1}b, 
the pump and probe microwave tones are combined and routed to the device through a circulator, which in turn directs the reflected signal to the measurement line. 
A calibrated, impedance-matched temperature-variable noise source~\cite{simbierowicz_characterizing_2021} is used to characterise the noise performance of the maser amplifier. 

The electron spins of the P1 centres are the active medium in our maser amplifier. 
These centres possess an electron spin $S=1/2$ and a nuclear spin $I=1$, with a hyperfine interaction tensor characterised by $\mathcal{A}_{\perp}/2\pi = 81.33 \,\text{MHz}$ and $\mathcal{A}_{||}/2\pi = 114.03 \,\text{MHz}$. 
Under a static magnetic field aligned parallel to the cubic crystallographic axis $[001]$ of the diamond, the electron spin states ($S = \pm 1/2$) split into three levels depending on the nuclear spin states, P1$_{+}$, P1$_{0}$, and P1$_{-}$, as shown in Fig.~\ref{fig:Fig2}a. 
The corresponding transition frequencies, denoted as $\omega_{+}$, $\omega_{0}$, and $\omega_{-}$, satisfy an energy conserving relationship $2\omega_{0} = \omega_{-} + \omega_{+}$, which facilitates a four-spin cross-relaxation process between the central transition ($m_{I}=0$) and the two satellite transitions ($m_{I}=\pm1$)~\cite{sorokin_cross_1960,ma_fourspin_2019,zhang_exceptional_2021}.

Maser amplification results from stimulated emission rather than wave-mixing processes, effectively acting as a phase-preserving negative-resistance amplifier.
Remarkably, noise entering the pump or ``idler-like'' modes does not contribute to the noise on the signal ~\cite{siegman_microwave_1964}. 
The operating principle of our maser amplifier is illustrated in Figs.~\ref{fig:Fig2}b–e. 
A static magnetic field $B_{\text{DC}}$ is applied to align the transition frequency $\omega_+$ with the resonator frequency $\omega_r$, while the P1$_0$ transition is pumped at $\omega_0/2\pi = 6.501\,\text{GHz}$ with a rate $\Gamma_\text{p}$ (Fig.~\ref{fig:Fig2}b). 
This pumping excites one P1$_+$ and one P1$_-$ spin by exchanging energy with two P1$_0$ spins through the four-spin cross-relaxation process~\cite{sorokin_cross_1960,ma_fourspin_2019} at a rate $\Gamma_{\text{cr}}$ (Fig.~\ref{fig:Fig2}c). 
Ordinarily, such pumping would saturate all transitions, but the dynamics are modified by the unequal relaxation times of the P1$_-$ and P1$_+$ spins (Fig.~\ref{fig:Fig2}f), with $T_{\text{1-}} \approx 1030\,\text{s}$ and $T_{\text{1+}} \approx 1650\,\text{s}$.
The faster relaxation of P1$_-$ is induced by another ``fast-relaxing'' spin system coexisting within the diamond, exhibiting $T_{\text{1}}$ of several tens of milliseconds (inset of Fig.~\ref{fig:Fig2}f). 
These spins have wide transition frequencies (Supplementary Information), mainly overlapping with the P1$_-$ transition at $\omega_{-}/2\pi = 6.407\,\text{GHz}$, thereby accelerating the relaxation of P1$_-$ through magnetic dipole–dipole interactions, as indicated by $\Gamma_{\text{P1}_{-}}^{\text{eff}}$ in Fig.~\ref{fig:Fig2}d.
This imbalance in relaxation times prevents saturation and sustains continuous excitation of the P1$_+$ state, resulting in population inversion. 
Although the microscopic origin of the fast-relaxing spins remains uncertain, they may be neutrally charged nitrogen-vacancy (NV$^{0}$) centres~\cite{baier_orbital_2020,barson_fine_2019} or nickel-related centres~\cite{isoya_epr_1990}. 
Charge-state conversion from NV$^-$ to NV$^0$ centres under optical pumping might offer further insight into their nature.
We refer to them as “fast-relaxing spins” throughout this paper.

\subsection*{Gain and bandwidth}

\begin{figure*}
    \centering
    \includegraphics[width=0.8\textwidth]{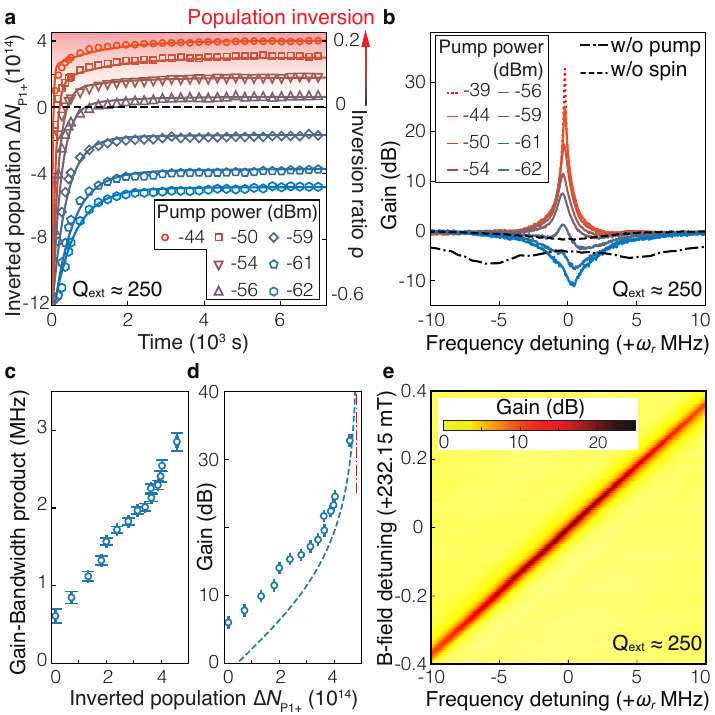}
    \caption{\textbf{Maser amplifier's gain and bandwidth.}
    \textbf{a}, Inverted spin population $\Delta N_\mathrm{P1_+}$ between the lower and upper P1$_{+}$ states over time under various pump powers for external quality factor of $Q_\text{ext} \approx 250$. 
    Open symbols represent the experimental data, while solid curves represent simulations. 
    \textbf{b}, Maser gain spectra under various pump powers. 
    The horizontal axis represents the detuning from the central resonator frequency $\omega_r/2\pi = 6.595\,\text{GHz}$. 
    The black dashed line represents the spectrum measured without the pump.
    \textbf{c}, Gain-bandwidth product versus inverted spin population. 
    \textbf{d}, Gain versus inverted spin population, where the open symbols and dashed curves are the experimental data and theoretical curves, respectively (Methods). 
    \textbf{e}, Maser gain spectrum versus magnetic field at a pump power of $P_\text{pump} = -44\, \text{dBm}$, yielding a gain of $23\,\text{dB}$. }
    
    \label{fig:Fig3}
\end{figure*}

Figure~\ref{fig:Fig3}a shows the time evolution of the inverted spin population $\Delta N_\mathrm{P1_+} = N_\mathrm{P1_+}^u - N_\mathrm{P1_+}^l$, between the upper ($u$) and lower ($l$) levels of P1$_+$ transition, extracted from the measured reflection coefficients (Methods). 
Above a pump power of $-58\,\text{dBm}$ at the device input, $\Delta N_\mathrm{P1_+}$ becomes positive, indicating population inversion. 
The solid curves show the simulated dynamics obtained from rate equations that include all the P1 states and their interactions (Methods). 
We also performed the same measurement with $Q_\mathrm{ext} \approx 730$, which clearly revealed the onset of self-oscillation (see Supplementary Information). 
Notably, superradiant emission~\cite{angerer_superradiant_2018,kersten_selfinduced_2026} 
cannot occur in the present system because the inversion builds up much more slowly than the dephasing time $T_2^* \sim 100$\,ns, leading to the loss of collective spin correlations.

Figure~\ref{fig:Fig3}b presents the maser gain spectra under various pump powers measured by a vector network analyser (VNA) under a static magnetic field of $B_\text{DC} = 232.15\,\text{mT}$. 
While the spectra show negative gains at weaker pump powers until $-59\, \text{dBm}$, the gain becomes positive above $-58\,\text{dBm}$. 
At a pump power of $-39\, \text{dBm}$, the gain exceeds $30\, \text{dB}$ with a bandwidth of approximately $70\,\text{kHz}$, which results in a gain-bandwidth product of $2.8\, \text{MHz}$. 
Figure~\ref{fig:Fig3}c shows the gain-bandwidth products versus inverted spin population $\Delta N_\mathrm{P1_+}$, 
whereas Fig.~\ref{fig:Fig3}d plots the gain as a function of $\Delta N_\mathrm{P1_+}$, with theoretical gain curves shown as dashed lines. 
While the gain–bandwidth product is limited to approximately $2.8\,$MHz, this could be overcome by employing multi-mode resonator architectures~\cite{sorokin_cross_1960} 
or a travelling-wave geometry~\cite{siegman_microwave_1964,glass_x-band_1994,chang_cavity_1959}.

Next, we demonstrate the frequency tunability of the diamond maser's gain profile by varying the external static magnetic field $B_{\text{DC}}$. 
As the transition frequencies of the P1 centres shift with $B_{\text{DC}}$, the maser's gain profile can also be tuned within the resonator bandwidth. 
Figure~\ref{fig:Fig3}e shows the measured maser gain versus $B_{\text{DC}}$, demonstrating tunability over $10\,\text{MHz}$ in the central gain and about $6\,\text{MHz}$ in the full-width at half-maximum. 

\subsection*{Maser noise}

\begin{figure*}
    \centering
    \includegraphics[width=\textwidth]{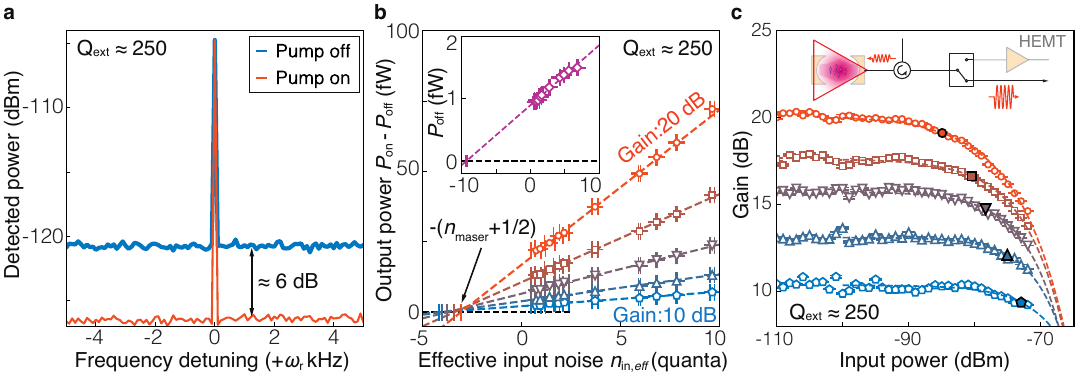}
    \caption{\textbf{Maser amplifier's noise and saturation power.} 
    \textbf{a}, Signal-to-noise ratio (SNR) in the presence of a weak coherent probe tone of $-140\,\text{dBm}$ with the maser amplifier turned on and off. 
    The `pump on' data, shown in red, is vertically shifted downwards to align the peak values for a clearer comparison of the SNRs. 
    The `pump off' data was measured with the spin ensemble detuned from the resonator (corresponding to the `w/o spin' condition in Fig.\ref{fig:Fig3}b). 
    The approximately $6\,\mathrm{dB}$ pump-on/off SNR change corresponds to approximately $4.4\,\mathrm{dB}$ after accounting for the $1.6\,\mathrm{dB}$ insertion loss. 
    \textbf{b}, Noise power $P_\mathrm{on} - P_\mathrm{off}$, detected by a spectrum analyser, as a function of the input noise $n_\text{in, eff}$ to the maser amplifier across various gains ($20$, $17.5$, $15$, $12.5$, and $10\, \text{dB}$). 
    Each $n_\text{in, eff}$ is corrected by taking the line attenuation and insertion losses into account (Methods). 
    The dashed lines are linear fits, whose intersection at $P_\text{on} - P_\mathrm{off} = 0$ determines the input-referred added noise of the maser amplifier~\cite{simbierowicz_characterizing_2021}. 
    The inset shows the system's background noise, dominated by the following HEMT amplifier, when the maser amplifier is not in operation. 
    \textbf{c}, Gain saturation measurement showing the maser amplifier gain as a function of input signal power. 
    The dashed curves are fits using a phenomenological model, $y(x)=a(1-ce^{b(x-x_0)})$, and the colored filled symbols denote the input $1\,\text{dB}$ compression points. 
    Inset: Schematic showing a switch to bypass the HEMT in order to prevent its saturation during these measurements.}
    
    \label{fig:Fig4}
\end{figure*}

To assess the noise performance of our maser amplifier, we measured the noise power spectrum in the presence of a weak microwave probe tone used as a reference, with the maser turned on and off. 
The results displayed in Fig.~\ref{fig:Fig4}a show a pump-on/off signal-to-noise ratio change of approximately $6\,\text{dB}$. 
After accounting for the measured insertion loss of $1.6\,\text{dB}$ (Extended Data Fig.1), this corresponds to an SNR improvement of approximately $4.4\,\text{dB}$. 
This indicates that the maser's added noise is about three times lower than that of the following measurement chain, which is dominated by the HEMT amplifier (noise temperature $\sim 2$--$3\,\mathrm{K}$). 

To quantitatively evaluate the maser amplifier's input-referred added noise, we used the impedance-matched temperature-variable noise source~\cite{simbierowicz_characterizing_2021}. 
This device enables independent control of the input noise $n_\text{in}$ (in units of photons), related to the noise source temperature $T_\text{in}$ (see Methods), allowing accurate determination of the amplifier's added noise. 
We then followed the analysis in Refs~\cite{day_roomtemperature_2024,parker_degenerate_2022}
to extract the added noise by the maser amplifier, $n_\mathrm{maser} + 0.5$, 
taking all the attenuation of the lines and insertion losses of the components into account (Methods). 

First, we characterised the baseline noise of our measurement line with the maser amplifier turned off. 
In this configuration, the noise power after the HEMT amplifier as a function of $n_{\mathrm{in, eff}}$, which is effective input noise (Methods), detected within a bandwidth of $\delta f_{\mathrm{BW}}$, is given by~\cite{simbierowicz_characterizing_2021,parker_degenerate_2022,day_roomtemperature_2024}
\begin{align}\label{Eq:main_HEMTnoiseTemp}
    \frac{P_\mathrm{off}}{\hbar\omega G_\mathrm{HEMT} \delta f_{\mathrm{BW}}} \approx n_{\mathrm{in, eff}} + \pab{n_\mathrm{HEMT} + \frac{1}{2}},
\end{align}
where $G_\mathrm{HEMT}$ and $n_{\mathrm{HEMT}}$ are the gain and added noise of the HEMT amplifier, respectively, and $\hbar$ is the reduced Planck constant. 
For simplicity, line attenuation and insertion losses are neglected in Eq.~\eqref{Eq:main_HEMTnoiseTemp}; the full expressions are provided in Methods. 
The results, shown in the inset of Fig.~\ref{fig:Fig4}b, indicate that $n_{\mathrm{HEMT}} \approx 9.33(86)$ quanta. 

Subsequently, we turned on the diamond maser amplifier and repeated the noise measurements at various maser amplifier gains. 
In this configuration, the difference in output noise power between the maser amplifier being on and off is given by~\cite{simbierowicz_characterizing_2021,parker_degenerate_2022,day_roomtemperature_2024}(Methods)
\begin{align}\label{eq:main_MaserNoiseTemp}    
    \frac{(P_\mathrm{on}-P_\mathrm{off})}{\hbar\omega G_\mathrm{maser} G_\mathrm{HEMT} \delta f_{\mathrm{BW}}} \approx n_{\mathrm{in, eff}} + \pab{n_\mathrm{maser} + \frac{1}{2}}.
\end{align}
Here, $G_\text{maser}$ and $n_\text{maser} + 0.5$ represent the gain and input-referred noise of the maser amplifier, respectively. 
As presented in Fig.~\ref{fig:Fig4}b, the minimum added noise of the maser amplifier is approximately $n_\mathrm{maser} +0.5 \approx 3.05(23)$ quanta, corresponding to an input-referred noise temperature of $T_\text{N} = (n_\mathrm{maser}+0.5)\,\hbar\omega_r/k_\mathrm{B} \approx 0.96\,\text{K}$.

Although the added noise is still higher than those typical of the superconducting parametric amplifiers, it mostly results from imperfect population inversion~\cite{siegman_microwave_1964,day_roomtemperature_2024}. 
While the spins in the higher energy level amplify the incoming signal through stimulated emission, those remaining in the lower energy level absorb photons. 
This counteraction effectively introduces additional noise from the spin ensemble, quantified by $n_\text{s} = \left(e^{\hbar \omega / k_B |T_s|} - 1 \right)^{-1}$, where negative spin temperature $T_{\text{s}}$ and the inversion ratio $\rho = \Delta N_{P1_{+}}/(N_\mathrm{P1_+}^u + N_\mathrm{P1_+}^l)$ are linked through 
\begin{equation}\label{Eq:inversionratio}
\rho = \tanh \left(\frac{\hbar \omega}{2 k_{\text{B}} |T_{\text{s}}|} \right), 
\end{equation}
with $k_\text{B}$ denoting the Boltzmann constant. 
From Eq.~\eqref{Eq:inversionratio}, the inversion ratio of $\rho \approx 0.17$ observed in Fig.~\ref{fig:Fig3}a (right-hand axis) at the gain of $20\,\text{dB}$ corresponds to a spin temperature of $|T_\text{s}| \approx 0.91\,\text{K}$, predicting an input-referred noise of about $3.29\,\text{quanta}$~\cite{day_roomtemperature_2024} (Methods, Eq.~\eqref{eq:sup_maser_added_noise}), in good agreement with the measured $n_{\text{maser}}$. 

It is important to note that the measured $n_\text{maser}$ values are by no means the fundamental limits in our system; they are constrained by the slow relaxation rate of the $\text{P1}_{-}$ spins, $\Gamma_{\text{P1}_{-}}^{\text{eff}} \approx 0.95$$\times 10^{-3}\,\text{s}^{-1}$, which in turn limits the effective pump rate on the P1$_{+}$ spins. 
With a slightly faster $\Gamma_{\text{P1}_{-}}^{\text{eff}}$ of $1.2 $$\times 10^{-3}\,\text{s}^{-1}$, which could be realized by engineering a higher density of the fast-relaxing spins, the noise could be reduced to $ 1.94\,\text{quanta}$ ($\rho\approx0.28$). 
Additionally, improving the internal quality factor $Q_\mathrm{int}$ by employing a high-$Q$ dielectric resonator~\cite{kato_highcooperativity_2023,hamamoto_dielectric_2024,breeze_continuouswave_2018,day_roomtemperature_2024,ng_quasicontinuous_2021,fahey_steadystate_2022}, 
with $Q_\mathrm{int} \sim 10^5$ achievable without compromising the coupling constant $g_0$, would further lower the loss-induced noise contribution by about $0.24$ quanta from $1.94$ to $1.70$ quanta. 
Exploring different spin systems and pump schemes could enable a near-perfect population inversion, bringing a quantum-limited maser amplifier within reach.

\subsection*{Compression point}
To assess the dynamic range of our maser amplifier, we measured the $1\,\text{dB}$ compression points at various maser amplifier gains. 
Figure~\ref{fig:Fig4}c shows the amplifier gains as a function of input probe power. 
The output (input) $1\,\text{dB}$ compression points ($P_\text{1dB}$) are measured to be approximately $-65 \, \text{dBm}$ ($-85 \, \text{dBm}$) at a gain of $20\, \text{dB}$, and approximately $-63\, \text{dBm}$ ($-73\, \text{dBm}$) at a gain of $10\, \text{dB}$. 
These measured compression points are $20$ to $30\, \text{dB}$ higher than those typically reported in Josephson travelling-wave parametric amplifiers~\cite{macklin_quantum_2015,esposito_perspective_2021}, but are two or three orders of magnitude lower than those achieved by state-of-the-art kinetic inductance travelling-wave parametric amplifiers~\cite{faramarzi_48_2024, parker_degenerate_2022}. 

\subsection*{Conclusion}

By revisiting masers at millikelvin temperatures, we demonstrated a near-quantum-limited diamond maser amplifier and achieved an input-referred noise of approximately $2.55$ quanta above the standard quantum limit. 
Although the noise, gain-bandwidth, and frequency-tunability are not yet comparable to those of state-of-the-art superconducting parametric amplifiers, these limitations are not fundamental and can be drastically improved by engineering the spin density, device geometry, or by exploring alternative spin systems and host materials. 
For operation as a practical millikelvin preamplifier, further reduction of insertion loss and pump power will also be required.
Maser amplification may emerge as a fundamentally distinct approach, operable under strong static magnetic fields 
(historically demonstrated up to several tesla\cite{momo_pulsed_1960a}) and potentially extendable to millimetre-wave frequencies (historically demonstrated up to W-band frequencies around $85$--$90\,\mathrm{GHz}$\cite{cardiasmenos_low_1976,sollner_lownoise_1979}), 
complementing related developments in kinetic-inductance-based amplifiers for magnetic-field-compatible~\cite{vaartjes_strong_2024, frasca_threewavemixing_2024, janssen_magneticfield_2025} and higher-frequency operations ~\cite{tan_operation_2024, kondo_predicted_2026, wang_kband_2026}.
Such a platform may be relevant to applications such as semiconducting spin-qubit readout, ultra-sensitive magnetic resonance spectroscopy, and fundamental physics experiments including dark-matter axion searches.
In the longer term, with magnetic-field shielding, miniaturisation, and broader operating bandwidth, the maser amplification mechanism may also find use within superconducting quantum systems.

\clearpage

\begin{acknowledgments}
We thank J. Pla, \c{C}.O. Girit, P. Bertet, A. Bienfait, D. Xiao, R.-B. Liu,  J.J.L. Morton, K. M{\o}lmer, E. Flurin, M. Hatifi, F. Quijandr\'{\i}a, J. Teufel, Y. Ota, S. Iwamoto, and S. Kono for their discussions, 
M. Ishida for assistance with device illustration, and D. Konstantinov and Y. Feng for providing access to dilution refrigerators in the early stages of this work.
We are also grateful to the members of the Hybrid Quantum Device Team within the Science and Technology Group and the Experimental Quantum Information Physics Unit at Okinawa Institute of Science and Technology (OIST) Graduate University for their insights. 
Additionally, we acknowledge the support from S. Ikemiyagi, M. Kuroda, and P. Kennedy at the Engineering Section of OIST. 
This work was supported by the Proof-of-Concept (PoC) program by OIST Innovation, the JST Moonshot R\&D Program (Grant No. JPMJMS2066), JST-PRESTO (Grant No. JPMJPR15P7), JSPS KAKENHI Grant-in-Aid for Scientific Research (B) (Grant No. 18H01817), Grant-in-Aid for Scientific Research on Innovative Areas (Grant No. 18H04295), the Science Research Promotion Fund from the Promotion and Mutual Aid Corporation for Private Schools of Japan (PMAC), and the Research Encouragement Grant from the AGC Inc. Research Foundation. 
M.O. and T.H. acknowledge financial support from the JSPS Grant-in-Aid for Fellows (Grant Nos. 23KJ2135 and 24KJ2178), and J.R.B. acknowledges support from OIST Graduate University.
\end{acknowledgments}

\section*{Author contributions}
M.O. and C.-P.L. contributed equally to this work and were primarily responsible for data collection with partial assistance from V.P.M.S., T.H., I.K., and Y.Ku. 
M.O. and C.-P.L. analysed the data with assistance from T.H., I.K., V.P.M.S., and Y.Ku. 
Y.Ku. conceived the project and initiated the preliminary experiments with J.R.B. and P.M. 
S.O., T.O., Y.Ko., and J.I. were responsible for the preparation of the diamond sample. 
Y.Ku. and H.T. jointly supervised the project. 
All authors contributed to the manuscript.

\section*{Competing interests}
The authors declare no competing interests.

\section*{Data availability}
The data presented in the figures, together with the analysis codes, have been deposited in the figshare repository at [DOI to be inserted upon deposition].

\clearpage

\bibliography{references}

\clearpage
\onecolumngrid

\setcounter{figure}{0}
\setcounter{table}{0}
\setcounter{section}{0}

\begin{center}
{\large\bfseries Methods and Supplementary Information for\\[4pt]
A near-quantum-limited diamond maser amplifier operating at millikelvin temperatures}
\end{center}

\vspace{1em}

\renewcommand{\figurename}{Extended Data Fig.}
\renewcommand{\tablename}{Extended Data Table}
\renewcommand{\thetable}{\arabic{table}}

\section*{Methods}

The experimental setup is described in detail in Supplementary Information Section 1.

\subsection*{Sample and device}

\subsubsection*{Diamond sample}
The diamond sample used in this work is identical to those described in previous works (Refs.~\cite{grezes_multimode_2014, ball_loopgap_2018}). 
It was synthesized by the temperature gradient method under high-pressure and high-temperature (HPHT) conditions of $5.5\,\text{GPa}$ and $1350\,^{\circ}\text{C}$ using an Fe-Ni-Co alloy as the metal solvent and natural diamond grains as the carbon source. 
A $(100)$ plate obtained by laser-cutting and polishing was irradiated with $3~\text{MeV}$ electrons at a fluence of $4\times 10^{17} \,\mathrm{e/cm^2}$ at room temperature, followed by annealing at $850~^{\circ}\text{C}$ for 2 hours in vacuum. 
Subsequently, it was further irradiated with $2\,\text{MeV}$ electrons at $700\,^{\circ}\text{C}$ to a fluence of $5\times 10^{17}\, \mathrm{e/cm^2}$ and annealed at $1000\,^{\circ}\text{C}$ for 2 hours in vacuum. 
The final dimensions of the diamond ($3 \times 1.5 \times 0.5 \,\mathrm{mm}^3$) used in this study were obtained from the central part of the plate via laser-cutting and polishing. 
Following electron irradiation and annealing, the concentrations of P1 centres and negatively-charged nitrogen-vacancy (NV$^-$) centres were measured to be approximately \(16\,\mathrm{ppm}\) and \(2\,\mathrm{ppm}\), respectively. 

\subsubsection*{Loop-gap resonator}

\begin{figure}[h]
    \centering
    \includegraphics[width=0.8\textwidth]{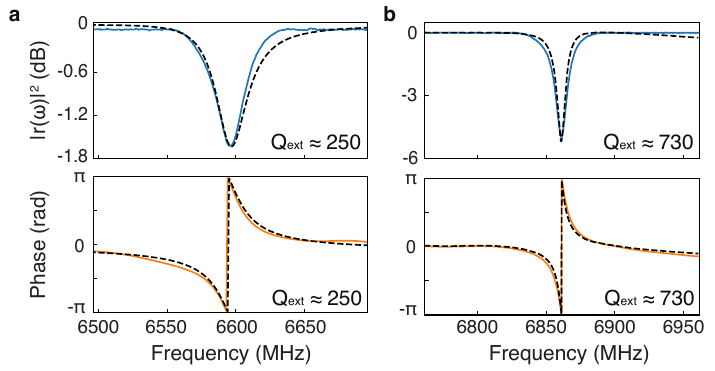}
    \caption{\textbf{Quality factors of the bare resonator.}
    Magnitude (upper panels) and phase (lower panels) of the resonator reflection coefficient $r(\omega)$ for (\textbf{a}) $Q_\mathrm{ext} \approx 250$ and (\textbf{b}) $Q_\mathrm{ext} \approx 730$. The black dashed lines represent the fits.}
    \label{figS:LGR}
\end{figure}

The microwave resonator used in this work is a loop-gap type~\cite{ball_loopgap_2018}. 
We determined its internal and external quality factors using the circle-fit method~\cite{probst_efficient_2015} after subtracting background amplitude oscillations to achieve a flat baseline. 
The results are shown in Extended Data Fig.~\ref{figS:LGR} for both $Q_\text{ext} \approx 250$ and $Q_\text{ext} \approx 730$. 
The internal quality factor $Q_\text{int}$ consistently stays around $3000$. 
The resonator frequency, $\omega_r$, is very sensitive to the distances of the two gaps. 
Consequently, after each disassembly and reassembly of the resonator and its enclosure to adjust $Q_\text{ext}$, 
$\omega_r$ varied over $6.59\,\text{GHz}$ to $6.86\,\text{GHz}$ upon each cooldown.

\subsection*{Maser noise measurement}\label{sec:sup_NoiseTemperature}

\subsubsection*{Amplification model}\label{subsubsec:AmplificationModel}

The propagation of signals through the measurement chain is modelled by combining the effects of amplification and attenuation.  
For a phase-insensitive amplifier with power gain $G$, the input–output relation is given by~\cite{caves_quantum_1982}
\begin{align}
    \hat{a}_\mathrm{out} = \sqrt{G}\,\hat{a}_\mathrm{in} + \sqrt{G-1}\,\hat{h}^\dagger,
\end{align}
where $\hat{a}_\mathrm{in}$ and $\hat{a}_\mathrm{out}$ denote the annihilation operators of the input and output modes, respectively, and $\hat{h}^\dagger$ represents an additional noise mode introduced by the amplification process.  
The corresponding mean photon number at the output is
\begin{align}
    n_\mathrm{out} = G \left(n_\mathrm{in}+\frac{1}{2}\right) + (G-1)\left(n_\mathrm{amp} + \frac{1}{2}\right),
\end{align}
with $n_\mathrm{in}$ the input photon number and $n_\mathrm{amp}$ the added noise of the amplifier.

\subsubsection*{Beam splitter model for loss}\label{subsubsec:BeamSplitterModel_for_inputnoise}

Losses in the signal path are modeled as an effective beam splitter with transmission coefficient $\beta$.  
In this case, the mean photon number at the output is expressed as~\cite{walls_quantum_2008}
\begin{align}\label{eq:sup_insertion_loss}
    n_\mathrm{out} = \beta \left(n_\mathrm{in}+\frac{1}{2}\right) + (1-\beta)\left(n_\mathrm{th} + \frac{1}{2}\right),
\end{align}
where $\beta$ represents the attenuation factor of the component, and $n_\mathrm{th}$ denotes thermal photon associated with a thermal bath at temperature $T$.

\subsubsection*{Y-factor method using a noise source}
We characterised the noise of the maser amplifier by using a procedure based on the Y-factor method with a calibrated noise source~\cite{simbierowicz_characterizing_2021}.  
The noise source is weakly thermalised to the cold plate at about $100\,\mathrm{mK}$, which allows for quasi-independent temperature control from $\approx 200\,\mathrm{mK}$ up to $\approx 5\,\mathrm{K}$ using an attached heater and thermometer. 
This setup enables precise control over the input noise temperature $T_\text{in}$ into the maser amplifier device. 

Thermal noise power is given by $P_\text{noise} = n \hbar \omega \delta f_\mathrm{BW}$, where $n$, $\hbar$, and $\delta f_\mathrm{BW}$ are the noise (in units of photons), the reduced Planck constant, and the measurement bandwidth, respectively. 
In our experiment, when the maser amplifier is turned off, $P_\text{off}$ can thus be expressed as:
\begin{align}\label{eq:sup_noisepower_HEMT}
    \frac{P_\text{off}(n_\mathrm{in})}{ \hbar\omega \delta f_\mathrm{BW}} 
    =& G_\text{HEMT} \beta_1 \beta_2 \left(n_\text{in} + \frac{1}{2}\right) 
    + G_\text{HEMT} (1-\beta_1)\beta_2 \left(n_0 + \frac{1}{2}\right) \nonumber\\
    &+ G_\mathrm{HEMT}(1-\beta_2) \left(n_0 + \frac{1}{2}\right)
    + (G_\mathrm{HEMT}-1) \left(n_\text{HEMT} + \frac{1}{2}\right),
\end{align}
where $G_\text{HEMT}$ and $n_\text{HEMT}$ represent the gain and added noise of the HEMT amplifier, respectively. 
$n_\mathrm{in}$ is related to the noise source temperature $T_\mathrm{in}$ as $n_\mathrm{in} = \left(e^{\hbar \omega / k_B T_\mathrm{in}} - 1 \right)^{-1}$.
$\beta_1$ and $\beta_2$ denote the attenuation from the noise source to the maser amplifier, and from the maser amplifier to the HEMT amplifier, respectively.
Here we ignore the additional noise contribution from the final room temperature amplifier, including cable attenuation from the HEMT to the room temperature setup, because Eq.~\eqref{eq:sup_noisepower_HEMT} can be approximated to Eq.~\eqref{Eq:main_HEMTnoiseTemp} in the main text, considering that $G_\text{HEMT} > 10^4$ and $T_\text{bkg} \approx 300\,\text{K}$. 

Similarly, the noise power with the maser amplifier on can be expressed as: 
\begin{align}
        \frac{P_\text{on} (n_\text{in})}{\hbar\omega \delta f_\mathrm{BW}} 
        =& G_\mathrm{maser}G_\text{HEMT} \beta_1 \beta_2 \left(n_\text{in} + \frac{1}{2}\right) 
        + G_\mathrm{maser}G_\text{HEMT} (1-\beta_1)\beta_2 \left(n_0 + \frac{1}{2}\right) \nonumber\\
        & + (G_\mathrm{maser}-1)G_\mathrm{HEMT}\beta_2\left(n_\mathrm{maser} + \frac{1}{2}\right)\nonumber\\
        &+ G_\mathrm{HEMT}(1-\beta_2) \left(n_0 + \frac{1}{2}\right) 
        + (G_\mathrm{HEMT}-1) \left(n_\text{HEMT} + \frac{1}{2}\right),\label{eq:sup_noisepower_maser}
\end{align}
where $G_\text{maser}$ and $n_\text{maser}$ represent the gain and added noise of the maser amplifier, respectively. 
Therefore, Eq.~\eqref{eq:sup_noisepower_maser} can be approximated to Eq.~\eqref{eq:main_MaserNoiseTemp} in the main text under the same considerations. 

Subtracting Eq.~\eqref{eq:sup_noisepower_HEMT} from Eq.~\eqref{eq:sup_noisepower_maser} yields
\begin{align}
    \frac{P_\text{on} - P_\mathrm{off}}{G_\mathrm{tot}\hbar\omega \delta f_\mathrm{BW}} 
    &= \beta_1 \pab{n_\mathrm{in} + \frac{1}{2}} 
    + (1-\beta_1) \pab{n_\mathrm{0} + \frac{1}{2}} + \pab{n_\mathrm{maser} + \frac{1}{2}} \\
    &= n_{\mathrm{in, eff}} + \pab{n_\mathrm{maser} + \frac{1}{2}}
\end{align}
where $G_\mathrm{tot}$ represents $G_\mathrm{tot} = (G_\mathrm{maser} -1) G_\mathrm{HEMT} \beta_2$, and $n_{\mathrm{in, eff}} =$ $ \beta_1 \pab{n_\mathrm{in} + 1/2}$ $ + (1-\beta_1) \pab{n_\mathrm{0} + 1/2}$ is the effective input noise entering the maser amplifier.

Experimentally, we sweep the number of input noise and measure $P_\mathrm{on}$ and $ P_\mathrm{off}$ at each point.
We estimate the condition $P_\mathrm{on} - P_\mathrm{off} = 0$ from measurement values, so the above equation becomes
\begin{align}
     n_{\mathrm{in, eff}}^\prime =  - \pab{n_\mathrm{maser} + \frac{1}{2}} 
\end{align}
where $n_{\mathrm{in, eff}}^\prime$ denote the intersection point of $P_\mathrm{on} - P_\mathrm{off}$ and $n_{\mathrm{in, eff}}$, which is negative value.

\subsubsection*{Noise of baseline}\label{sec:sup_baseline_noise}

\begin{figure}[h]
    \centering
    \includegraphics[width=0.8\textwidth]{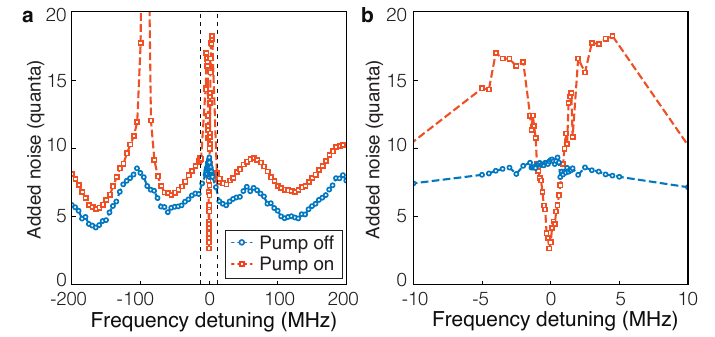}
    \caption{\textbf{Noise measurements over wide frequency range.}
    Added noise as a function of frequency detuning, measured with the pump on (red squares), and with the pump off and an off-resonant magnetic field for P1$_+$ (blue circles). 
    \textbf{a}, Wide range: $\pm 200\, \mathrm{MHz}$. 
    \textbf{b}, Zoomed-in view: $\pm 10\, \mathrm{MHz}$, corresponding to the dashed window in \textbf{a}.}
    \label{figS:baseline_noisetemp}
\end{figure}

We measured the added noise at several frequencies around the maser peak gain, both with the pump on and off, as shown in Extended Data Fig.~\ref{figS:baseline_noisetemp}.
First, we measured the added noise with the pump off, namely, $n_\text{HEMT}$. 
To minimize the absorption effect by the spins, which is equivalent to having additional attenuation along the line, the magnetic field was detuned from the resonance of P1$_+$. 

Under these conditions, as seen in Extended Data Fig.~\ref{figS:baseline_noisetemp}, the added noise is measured to be between $4$ and $10\, \mathrm{quanta}$, in reasonable agreement with the values reported in the HEMT's datasheet. 
The oscillations are mostly attributed to small impedance mismatches among the components along the measurement chain from $10\,\text{mK}$ up to the HEMT amplifier. 
At the resonance frequency $\omega_r \approx 6.6\, \text{GHz}$, it reaches $\approx 9.3\, \text{quanta}$, which is used as the HEMT's noise photon $n_\text{HEMT}$ in the main text. 
During this measurement with the pump off, we disconnected both the pump and cancellation lines and terminated at room temperature to minimize the noise from the room-temperature setup, especially from the amplifier in the cancellation line (see Fig.~S2). 

When the pump and cancellation tones are turned on and tuned for maser amplifier's gain of $20\, \mathrm{dB}$, the noise baseline increases by approximately $2\,\text{quanta}$ over the entire frequency range (except near $\omega_r$, i.e., the maser amplifier frequency), most likely due to noise coming from the room-temperature amplifier connected to the ``Cancellation'' line. 
The significant increase of the noise around $\sim -100\, \mathrm{MHz}$ is due to the pump.

\subsubsection*{Detailed line attenuation characterization}\label{sec:sup_attenuation_meas}
\begin{figure}[h]
    \centering
    \includegraphics[width=0.8\textwidth]{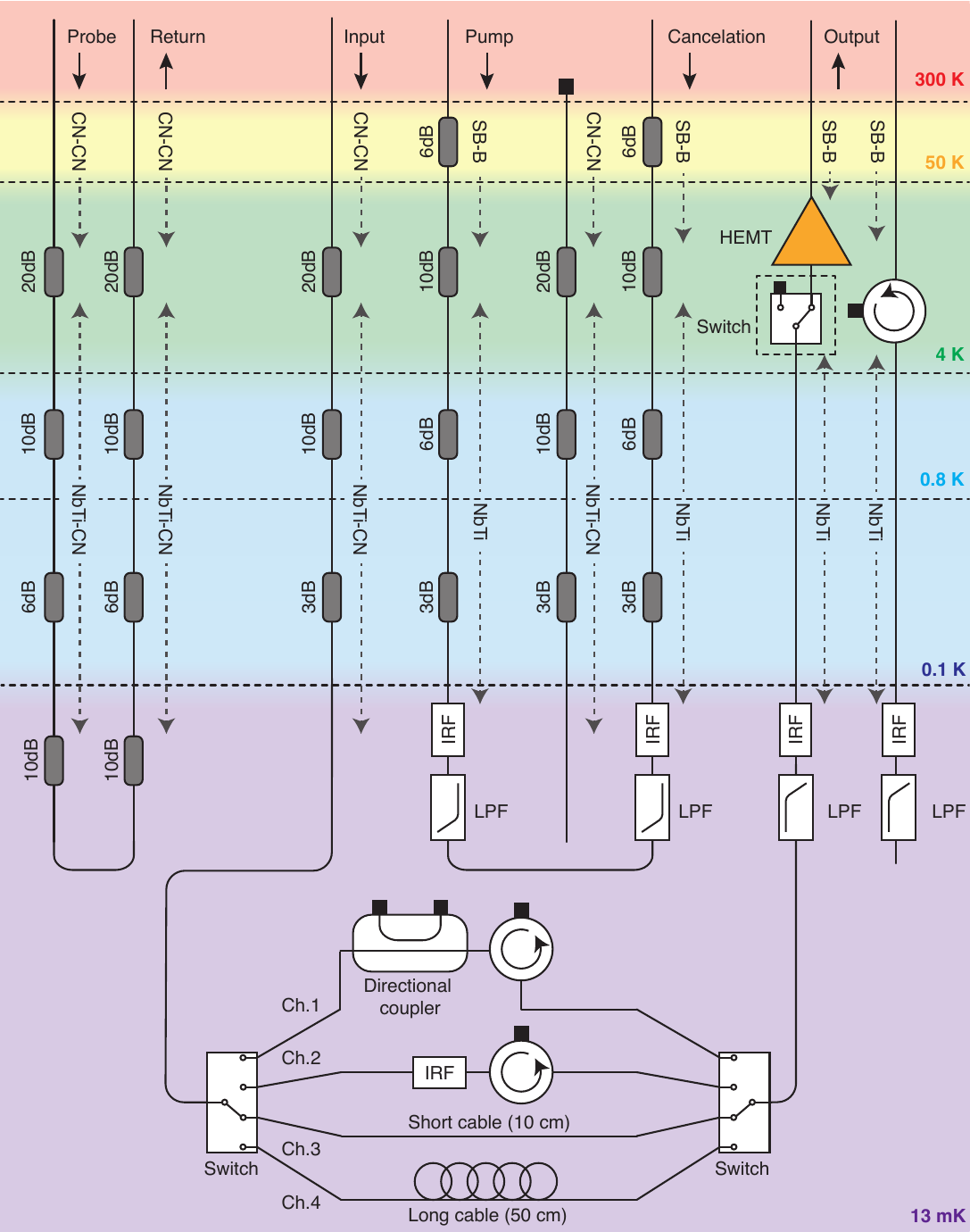}
    \caption{\textbf{Cryogenic wiring used to characterise the line attenuations.} 
         The vertical arrows next to each line denote the materials used for the inner and outer conductors, with CN, B, and SB representing cupronickel, beryllium copper, and silver-plated beryllium copper, respectively. 
             }
    \label{figS:Cryo_setup_lineatts}
\end{figure}

In order to accurately estimate the power of the pump and probe tones arriving at the input of the maser amplifier device, we independently characterised the attenuation of both the probe and pump lines in a separate cooldown. 
To this end, we measured the transmission coefficients over the ``symmetric lines'', i.e, pairs of lines with identical configurations, including coaxial cable materials, lengths, and attenuators at each temperature stage, as illustrated in Extended Data Fig.~\ref{figS:Cryo_setup_lineatts}. 
These symmetric lines were interconnected by a short, low-loss coaxial cable at the mixing chamber stage. 

We also characterised the attenuation of the cables and components in the Probe line at the mixing chamber stage, specifically from the infrared filter through the circulator, noise source, directional coupler, and the final circulator to the input of the maser device (see also Supplementary Information Section 1.1). This was performed using two mechanical microwave switches (Radiall, R573423600), as depicted in Extended Data Fig.~\ref{figS:Cryo_setup_lineatts}. 
Four measurement paths (Ch.~1–Ch.~4) were established for this purpose. 
Channels 1 and 2 correspond to the paths from the noise source to the maser amplifier device and from the infrared filter to the noise source, respectively. 
Channel 3 was used to isolate the attenuation introduced by the components measured in Channels 1 and 2 by using a low-loss, flexible coaxial cable.
Channel 4 was used to estimate the attenuation of the low-loss flexible coaxial cables used to connect the components by comparing with Channel 3. 
The results of these measurements are summarized in Table~\ref{tab:line_atts}. 
Note that the insertion loss of the mechanical switches must also be taken into account in the calculation of Eq.~\eqref{eq:sup_insertion_loss} for the noise measurement; here we relied on the value on the datasheet of $0.3\,\text{dB}$.

\subsection*{Extracting spin information through resonator}

\subsubsection*{Single Spin Coupling Constant}

The single-spin coupling constant $g_0$, which represents the interaction strength between a single P1 centre and a single microwave photon in a resonator, is expressed as~\cite{kubo_strong_2010}
\begin{align}
    g_0 = - \gamma_e \delta B_0 \left\langle -\tfrac{1}{2} \middle| \mathbf{S} \middle| +\tfrac{1}{2} \right\rangle,
\end{align}
where $\gamma_e$ and $\delta B_0$ denote the gyromagnetic ratio of the electron spin and the magnetic vacuum fluctuation of the resonator, respectively, and $\mathbf{S}$ is the spin vector operator with $|{-\frac{1}{2}}\rangle$ and $|{+\frac{1}{2}}\rangle$ are the P1 centre's electron spin eigenstates. 
The magnetic vacuum fluctuation is given by
\begin{align}
    \delta B_0 = \sqrt{\frac{\mu_0 \hbar \omega_r}{2 \mathcal{V}_\mathrm{eff}}},
\end{align}
where $\mu_0$, $\omega_r$, and $\mathcal{V}_\mathrm{eff}$ are the vacuum permeability, the resonator frequency, and the effective mode volume of the resonator, respectively. 
$\mathcal{V}_\mathrm{eff}$ is defined as the ratio of the integral of the magnetic field energy density over the whole resonator mode volume $V$ to the maximum magnetic energy density~\cite{choi_ultrastrong_2023}
\begin{align}
    \mathcal{V}_\mathrm{eff} = \frac{\int_V |\mathbf{B}(\mathbf{r})|^2 /\mu(\mathbf{r}) \mathrm{d}V}{ |\mathbf{B}(\mathbf{r}_0)|^2 / \mu(\mathbf{r}_0)},
\end{align}
where $\mathbf{B}(\mathbf{r})$ is the magnetic field in the resonator, maximized at the position $\mathbf{r}_0$. 
Using a finite element simulation software, COMSOL Multiphysics, the effective resonator mode volume $\mathcal{V}_\mathrm{eff}$ of our loop-gap resonator is estimated to be approximately $\mathcal{V}_\mathrm{eff} \approx 2.5 \times 10^{-8}\, \mathrm{m^3}$. 
Therefore, the single-spin coupling constant of the P1 centre with $S=\frac{1}{2}$ is obtained to be $g_0 \approx 2\pi \times 0.15\, \mathrm{Hz}$. 

\subsubsection*{Input-output theory}
To determine the number of inverted or non-inverted spins through the resonator reflection \( r(\omega) \), we employed the input-output theory~\cite{collett_squeezing_1984} for a coupled resonator-spin ensemble system with a coherent probe tone \(\alpha_{\text{in}} = \alpha_0 e^{-i \omega t}\)~\cite{grezes_multimode_2014,day_roomtemperature_2024}.
The Hamiltonian can be expressed using the Tavis-Cummings model with Holstein-Primakoff approximation~\cite{diniz_strongly_2011,kurucz_spectroscopic_2011,julsgaard_dynamical_2012} as follows:
\begin{align}
    H / \hbar = \omega_r a^\dag a + \frac{1}{2} \sum_{j=1}^N \omega_j s^\dag_j s_j + \sum_{j=1}^N g_j \left( s^\dag_j a + s_j a^\dagger \right) - i \sqrt{\kappa_\mathrm{ext}} (\alpha_\text{in} a^\dag - \alpha_\text{in}^\ast a),
\end{align}
where \(a\) (\(s_j\)) and \(a^\dagger\) (\(s_j^\dagger\)) denote the annihilation and creation operators for the resonator mode (\(j\)-th spin), respectively. 
In addition, $N$ represents the number of spins, and $g_j$ is the single spin coupling constant between the resonator and the $j$-th spin. 
\(\kappa_{\mathrm{ext}} = \omega_r / Q_{\mathrm{ext}}\) and \(\kappa_\mathrm{int} = {\omega_r}/{Q_\mathrm{int}}\) are the external and internal loss rates of the resonator, with \(Q_\mathrm{ext}\) (\(Q_\mathrm{int}\)) representing the external (internal) quality factor. 

\subsubsection*{The spin function $K(\omega)$}
Following the standard method for deriving the reflection coefficient~\cite{collett_squeezing_1984}, we obtain
\begin{align}
    r(\omega) = \frac{i\kappa_\mathrm{ext}}{(\omega - \omega_r) + i\frac{\kappa_\mathrm{ext} + \kappa_\mathrm{int}}{2} \pm K(\omega)} -1, \label{eq:sup_reflection}
\end{align}
where $K(\omega)$, called the ``spin function'', includes all information about the spin ensemble. 
As will be discussed in the following section, $K(\omega)$ is linked to the spin susceptibility~\cite{grezes_multimode_2014} and is defined as shown in Refs.~\cite{diniz_strongly_2011,kurucz_spectroscopic_2011,julsgaard_dynamical_2012},
\begin{align}
    K(\omega) = \sum_{j=1}^{N} \frac{g_j^2}{(\omega - \omega_j) + i \gamma_j /2} = \frac{g_\text{ens}^2}{(\omega - \omega_s) + i\Gamma/2}, 
\end{align}
with \(g_\text{ens} = \sum_j g_j = g_0 \sqrt{N}\) representing the ensemble coupling constant between the resonator and the spin ensemble. 
The decay rate of individual spins is denoted by \(\gamma_j\), while \(\omega_s\) and \(\Gamma/2\) represent the centre frequency and the half-width at half-maximum (HWHM) of the spin ensemble, respectively, assumed to follow a Lorentzian distribution here. 
The sign in front of $K(\omega)$ in Eq.~\eqref{eq:sup_reflection} represents the state of the spin ensemble, i.e., a positive sign ($+$) corresponds to an inverted spin ensemble, whereas a negative sign ($-$) represents a non-inverted spin ensemble. 

We modify $K(\omega)$ to incorporate the spin polarization $\bar{\rho} = \Delta N / N_\mathrm{tot}$, where $\Delta N = N_l - N_u$ represents the population difference between the lower and upper levels of the $\text{P1}_{+}$ state, and $N_\mathrm{tot} = N_l + N_u$ is the total number of spins in the $\text{P1}_{+}$ transition. 
In a system with a population difference $\Delta N$, the ensemble coupling constant $g_\text{ens}$ is replaced by the effective coupling constant $g_\text{ens,eff}$, defined as $g_\text{ens,eff}^2 = g_0^2 \Delta N$. 
Consequently, $K(\omega)$ is reformulated in terms of $\bar{\rho}$ as
\begin{align}
    K(\omega) = \frac{g_\text{ens,eff}^2 }{(\omega - \omega_s) + i\Gamma/2} = \frac{g_0^2 \Delta N}{(\omega - \omega_s) + i\Gamma/2} = \frac{\bar{\rho} g_\text{ens,total}^{2}}{(\omega - \omega_s) + i\Gamma/2}, \label{eq:sup_Kw}
\end{align}
where $g_\mathrm{ens,total}^2 = g_0^2 N_\mathrm{total}$ represents the ``conventional'' collective coupling constant, assuming all spins are polarized into the lower energy state. 
The sign of $K(\omega)$ reflects the sign of $\Delta N$; it becomes negative in the case of population inversion and positive in a non-inverted state. 
To ensure consistency with Eq.~\eqref{eq:sup_reflection}, we rewrite the reflection coefficient $r(\omega)$ as follows 
\begin{align}
    r(\omega) = \frac{i\kappa_\mathrm{ext}}{(\omega - \omega_r) + i\frac{\kappa_\mathrm{ext} + \kappa_\mathrm{int}}{2} - K(\omega)} -1. \label{eq:sup_ref_Kw}
\end{align}

\subsubsection*{Measurement of population differences $\Delta N$}\label{subsubsec:DeltaN_from_reflection}
\begin{figure}[h]
    \centering
    \includegraphics[width=0.8\textwidth]{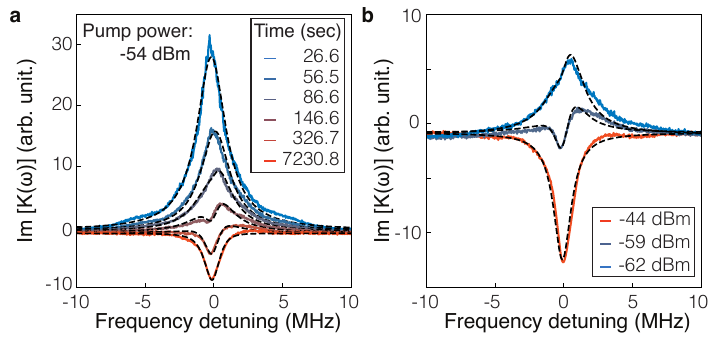}
    \caption{\textbf{Measured spin function $K(\omega)$.}
    Imaginary parts of $K(\omega)$ obtained from the measured maser gain spectra, with fits (black dashed lines) with Eq.~\eqref{eq:sup_dbl_lorentzian}. 
    The data shown in \textbf{a} correspond to selected time points from the time dynamics presented in Fig.~\ref{fig:Fig3}a in the main text at a pump power of $-54\, \mathrm{dBm}$, while those in \textbf{b} represent gain spectra under the three pump powers after the gain was stabilized. }
    \label{figS:Kmat}
\end{figure}

The function $K(\omega)$ can be expressed in terms of the reflection coefficient $r(\omega)$ by rearranging Eq.~\eqref{eq:sup_ref_Kw}: 
\begin{align}\label{eq:sup_K_of_reflection}
    K(\omega) = (\omega - \omega_r) + \frac{i}{2} \left( \kappa_\mathrm{int} + \kappa_\mathrm{ext} \frac{r(\omega) - 1}{r(\omega) +1} \right), 
\end{align}
which allows $K(\omega)$ to be directly determined from the experimentally obtained $r(\omega)$ data. 
Extended Data Fig.~\ref{figS:Kmat} shows such examples converted from experimentally obtained data and fits using the imaginary parts of Eq.~\eqref{eq:sup_Kw}.

In an ideal spin ensemble, the imaginary part of $K(\omega)$ exhibits a Lorentzian profile, enabling straightforward determination of the population difference $\Delta N$ by fitting the measured spectrum using the function: 
\begin{align}
    y = y_0 + \frac{A}{(x-x_0)^2 + B^2}.
\end{align}
From this function and the imaginary part of Eq.~\eqref{eq:sup_Kw}, the population difference $\Delta N$ can be extracted as $\Delta N = A/g_0^2B$.

However, significant deviations from the ideal Lorentzian profile are observed under certain experimental conditions. 
Immediately after initiating the pump, and before reaching the steady state, the spectrum becomes distorted due to the rapid inversion of the spins near the centre of the distribution, while those farther from the centre remain non-inverted. 
Furthermore, even in the steady state, intermediate pump powers lead to partial inversion of the spin ensemble, resulting in non-uniform inversion and persistent spectral distortion. 
This behavior is illustrated in Extended Data Fig.~\ref{figS:Kmat}. 
Panel a shows spectra of pump power $-54\,\mathrm{dBm}$ at selected time points extracted from the time evolution in Fig.~\ref{fig:Fig3}a of the main text. 
Among these, the spectra at 146.6 and 326.7 seconds clearly exhibit transient distortions following pump initiation. 
Panel b shows spectra at various pump powers extracted from the gain profiles in Fig.~\ref{fig:Fig3}a after gain stabilization. 
Notably, the spectrum at a pump power of $-59\,\mathrm{dBm}$ exhibits pronounced steady-state distortion due to partial inversion. 

To accurately account for these spectral conditions, we employ a two-Lorentzian fitting model in which the spectrum is represented as a superposition of two separate Lorentzian contributions; one representing the inverted upper-level spins and the other the lower-level spins:
\begin{align}
    y = y_0 + \frac{A_1}{(x-x_1)^2 + B_1^2} - \frac{A_2}{(x-x_2)^2 + B_2^2}. \label{eq:sup_dbl_lorentzian}
\end{align}
The results of fits using this model are illustrated in Extended Data Fig.~\ref{figS:Kmat}.
From this fitting model, the population difference $\Delta N$ can be described as 
\begin{align}
    \Delta N = \frac{1}{g_0^2}\left(\frac{A_2}{B_2} - \frac{A_1}{B_1} \right) .
\end{align}

\subsubsection*{Maser gain}\label{sec:masergain}
The peak maser gain is defined by the absolute square of the reflection coefficient $r(\omega)$ at the resonance, where $\omega = \omega_r = \omega_s$. 
According to Eq. \eqref{eq:sup_ref_Kw}, this relationship can be expressed as
\begin{align}\label{Eq:gain}
    G = |r(\omega)|^2 = \frac{(\kappa_\mathrm{ext} - \kappa_\mathrm{int} + |\kappa_{s}|)^2}{(\kappa_\mathrm{ext} + \kappa_\mathrm{int} - |\kappa_{s}|)^2},
\end{align}
where we defined a new parameter, $\kappa_s = 4g_\mathrm{ens}^2/\Gamma = \mathrm{Im} [2K(\omega_r = \omega_s)]$~\cite{day_roomtemperature_2024, julsgaard_dynamical_2012}, as the ``spin transition rate''. 
This parameter quantifies the effective transition rate of the spin ensemble to the resonator mode, i.e., the rate at which spins undergo upwards (downwards) transitions per unit time through the stimulated emission (absorption) with a rate $\Gamma_\text{s} = 4g_{0}^{2}/\Gamma$. 
Moreover, $\kappa_\text{s}$ is directly correlated with the ``magnetic quality factor'' $Q_\text{mag}$~\cite{siegman_microwave_1964, sherman_diamondbased_2022}, which characterises a spin ensemble system as a maser gain medium: 
\begin{align}
    Q_\text{mag} = \frac{1}{\eta \chi^{\prime\prime}} = \frac{1}{\eta\cdot \textrm{Im} [2K(\omega_r=\omega_s)/(\omega_r\eta)]} = \frac{\omega_r \Gamma}{4 g_\mathrm{ens}^2} = \frac{\omega_r}{\kappa_s},
\end{align}
where $\eta$ is the magnetic field filling factor and $\chi^{\prime\prime}$ is the imaginary part of the spin susceptibility at resonance, defined as $\chi(\omega) = -2 K(\omega) / (\omega_r \eta)$~\cite{grezes_multimode_2014}. 

Using Eq.~\eqref{Eq:gain}, we calculated the gains $G (\omega_r = \omega_s)$ for the two external quality factors, $Q_\text{ext} \approx 250$ and $730$, used in the experiments. 
The results are shown as dashed curves in Fig.~\ref{fig:Fig3}d. 

\subsubsection*{Added noise of the maser amplifier}\label{sec:sup_maser_added_noise}
Following Ref.~\cite{day_roomtemperature_2024} (Eqs.~B48--B52 therein), the input-referred noise of the maser amplifier is the sum of the noise from the inverted spin ensemble and the thermal noise entering through the internal loss of the resonator. 
With a sufficiently high maser gain and on resonance, $|\kappa_{s}| \lesssim \kappa_\mathrm{ext} + \kappa_\mathrm{int}$ in Eq.~\eqref{Eq:gain}, this reduces to
\begin{align}\label{eq:sup_maser_added_noise}
    n_\mathrm{maser}^\mathrm{model} \approx \pab{1 + \frac{\kappa_\mathrm{int}}{\kappa_\mathrm{ext}}} \pab{n_\mathrm{s} + \frac{1}{2}} + \frac{\kappa_\mathrm{int}}{\kappa_\mathrm{ext}}\pab{n_\mathrm{th} + \frac{1}{2}},
\end{align}
where $n_\mathrm{s} = \left(e^{\hbar\omega/k_\mathrm{B}|T_\mathrm{s}|} - 1\right)^{-1}$ is the spin noise set by the inversion ratio $\rho$ through Eq.~\eqref{Eq:inversionratio}, and $n_\mathrm{th}$ is the thermal photon of the resonator bath. 
For $\rho \approx 0.17$ and $\kappa_\mathrm{int}/\kappa_\mathrm{ext} = Q_\mathrm{ext}/Q_\mathrm{int} \approx 0.1$ at millikelvin temperatures ($n_\mathrm{th} \approx 0$), Eq.~\eqref{eq:sup_maser_added_noise} gives $n_\mathrm{maser}^\mathrm{model} \approx 3.3$ quanta, in agreement with the measured $3.05(23)$ quanta. 
In the limit $\rho \to 1$ and $\kappa_\mathrm{int}/\kappa_\mathrm{ext} \to 0$, it reduces to the standard quantum limit of half a quantum.
Note that $n_\mathrm{maser}^\mathrm{model}$ includes the quantum-noise contribution, thus $n_\mathrm{maser}^\mathrm{model} = n_\mathrm{maser} + 1/2$.

\subsection*{Modeling spin dynamics}

\subsubsection*{Semi-classical Rate Equations}

We modelled the dynamics of the diamond maser amplifier using semiclassical rate equations~\cite{sorokin_cross_1960}. 
Although the quantum-Langevin equations can, in principle, yield equivalent dynamics ~\cite{jin_proposal_2015,day_roomtemperature_2024}, we opted for the semiclassical approach for numerical simplicity. 
Furthermore, while the quantum-Langevin equations may in principle describe the noise, their applicability to the present maser amplifier system remains to be verified; this issue is under investigation and will be reported elsewhere. 

The system under consideration, as depicted in Extended Data Fig.~\ref{figS:rate_eqs}a, consists of eight energy levels, which include two levels of the ``fast-relaxing spins'' and the three two-level systems associated with the P1 centres nuclear spin states, P1$_+$, P1$_0$, and P1$_-$. 
This configuration leads to eight rate equations, in addition to a photon number rate equation, to describe the population dynamics and photon numbers within the system. 
However, due to the computational complexity of solving nine equations, we simplified the model to seven equations by combining the two levels of $\text{P}1_{-}$ with the fast-relaxing spins. 
The resulting set of seven rate equations, including the photon number rate equation, are represented as follows~\cite{sorokin_cross_1960, carmichael_statistical_2013}:
\begin{align}
    \dot{N_1} &= \Gamma_{41} N_4 + \frac{\Gamma_\mathrm{cr}}{N_\text{tot}^3} (N_2^2 N_4 N_6 - N_5^2 N_1 N_3) - \Gamma_s n (N_1 - N_4), \label{eq:sup_rateeqs}\\
    \dot{N_2} &= \Gamma_{52} N_5 - 2\frac{\Gamma_\mathrm{cr}}{N_\text{tot}^3} (N_2^2 N_4 N_6 - N_5^2 N_1 N_3) - \Gamma_\text{p} (N_2 - N_5), \nonumber\\
    \dot{N_3} &= \Gamma_{63}^\mathrm{eff} N_6 + \frac{\Gamma_\mathrm{cr}}{N_\text{tot}^3} (N_2^2 N_4 N_6 - N_5^2 N_1 N_3), \nonumber\\
    \dot{N_4} &= -\Gamma_{41} N_4 - \frac{\Gamma_\mathrm{cr}}{N_\text{tot}^3} (N_2^2 N_4 N_6 - N_5^2 N_1 N_3) + \Gamma_s n (N_1 - N_4), \nonumber\\
    \dot{N_5} &= -\Gamma_{52} N_5 + 2\frac{\Gamma_\mathrm{cr}}{N_\text{tot}^3} (N_2^2 N_4 N_6 - N_5^2 N_1 N_3) + \Gamma_\text{p} (N_2 - N_5) , \nonumber\\
    \dot{N_6} &= -\Gamma_{63}^\mathrm{eff} N_6 - \frac{\Gamma_\mathrm{cr}}{N_\text{tot}^3} (N_2^2 N_4 N_6 - N_5^2 N_1 N_3), \nonumber
\end{align}
where $N_i$ represents the number of spins of each level, $\Gamma_{41}$, $\Gamma_{52}$, and $\Gamma_{63}$ are the relaxation rates of each transition, $\Gamma_\mathrm{cr}$ is the four-spin cross-relaxation rate, and $\Gamma_s = 4g_0^2/\Gamma$ is the stimulated emission rate. 
The rate equation for the intra-resonator photon number $n$ is: 
\begin{align}
\dot{n} &= -\kappa_\mathrm{tot} n - \Gamma_s n (N_1 - N_4) + \Gamma_s N_4. \label{eq:sup_photon_eq}
\end{align}

\subsubsection*{Cross-relaxation rate}

The cross-relaxation rate $\Gamma_\mathrm{cr}$ in our system cannot be directly measured due to the presence of the ``fast-relaxing spins'', contrasting with the previous reports~\cite{sorokin_cross_1960,ma_fourspin_2019,zhang_exceptional_2021}. 
To estimate this rate indirectly, we compared computational results from the rate equations with experimental results obtained under various pump powers, as depicted in Extended Data Fig.~\ref{figS:rate_eqs}b and explained in the next subsection. 
We determined $\Gamma_\mathrm{cr}$ to be within the range of $\sim 1$ to $10\,\text{mHz}$, with the parameters listed in Tables~\ref{tab:relax_rates} and \ref{tab:rate_eqs_params}. 
While previously reported cross-relaxation rates for highly nitrogen-doped diamonds typically range from $\sim 10\,\text{Hz}$ to $\sim 100\,\text{Hz}$~\cite{sorokin_cross_1960, ma_fourspin_2019}, theoretical models of the four-spin interactions imply that $\Gamma_\mathrm{cr} \propto d^6$~\cite{ma_fourspin_2019}, where $d$ denotes the spin density. 
This supports our findings of $\Gamma_\mathrm{cr} \sim \text{mHz}$, as the P1 centre density in our diamond sample of $\approx 16\, \text{ppm}$ is considerably lower than the $> 100\,\text{ppm}$ in those previous studies. 

\subsubsection*{Calculation of pump rate $\Gamma_\text{p}$}\label{subsubsec:pumprate}
Spins absorb incoming microwave pump tones and are excited to the upper state, a process governed by the same underlying physics as stimulated emission. 
Therefore, the photon absorption rate, i.e., the pump rate $\Gamma_\text{p}$, is calculated as a product of the stimulated \textit{absorption} rate and the number of incoming photons, as described in Eqs. (7.7) and (7.8) on p.260 in Ref.~\cite{carmichael_statistical_2013}:
\begin{align}
\Gamma_\text{p} = \frac{g_0^2 \Gamma}{(\omega_p - \omega_s)^2 + \Gamma^2/4} n_\mathrm{in},
\end{align}
where $\omega_p$ is the pump frequency, $\omega_s$ is the average spin resonance frequency of P1$_0$, and $n_\mathrm{in}$ is the number of input photons. 
The expression ${g_0^2 \Gamma}/\left[{(\omega_p - \omega_s)^2 + \Gamma^2/4}\right]$ on the right-hand side is the rigorous stimulated emission (absorption) rate with detuning between $\omega_p$ and $\omega_s$.

\begin{figure}[h]
    \centering
    \includegraphics[width=0.8\textwidth]{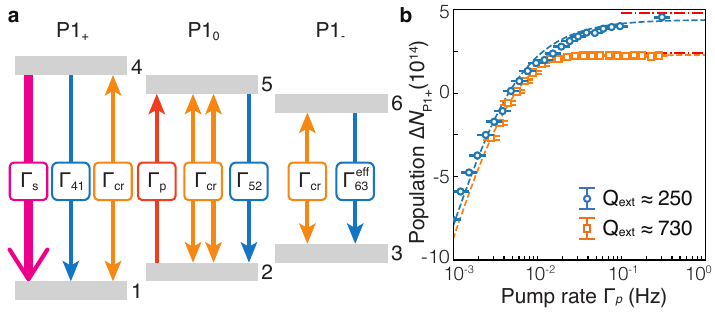}
    \caption{\textbf{Rate equation model and comparison to measurements.}
    \textbf{a}, Energy level diagram of P1 centre and the relaxation rates used in the rate-equation model. 
    \textbf{b}, Steady-state population difference $\Delta N_\mathrm{P1_+}$ between the lower and upper P1$_+$ states, obtained from measured $r(\omega)$ using Eq. \eqref{eq:sup_K_of_reflection} in the Materials and Methods, plotted versus the pump rate $\Gamma_\mathrm{p}$ calculated from Eq.~\eqref{eq:sup_pump_freq} based on the applied pump power $P_\mathrm{in}$. 
    The dashed lines indicate the results of rate-equation simulations, and the red dot-dashed lines represent the threshold of both $Q_\mathrm{ext}$ values.}
    \label{figS:rate_eqs}
\end{figure}

From the input-output theory, $n_\mathrm{in}$ with a coherent pump tone $\beta_\mathrm{in} = \beta_0 e^{-i\omega t}$ applied to a single-port resonator is expressed as
\begin{align}
    n_\mathrm{in} &= \frac{4\kappa_\mathrm{ext}}{\kappa_\mathrm{tot}^2 + 4(\omega_p - \omega_r)^2} |\beta_0|^2 = \frac{4\kappa_\mathrm{ext}}{\kappa_\mathrm{tot}^2 + 4(\omega_p - \omega_r)^2} \frac{P_\mathrm{in}}{\hbar\omega},
\end{align}
where $P_\mathrm{in} = \hbar\omega |\beta_0|^2$ denotes the input pump power, and $\kappa_\mathrm{tot} = \kappa_\mathrm{ext} + \kappa_\mathrm{int}$ is the total loss rate of the resonator.
Therefore, the pump rate $\Gamma_\text{p}$ can be expressed as a function of $P_\mathrm{in}$ as follows;
\begin{align}
    \Gamma_\text{p} (P_\mathrm{in}) = \frac{g_0^2 \Gamma}{(\omega_p - \omega_s)^2 + \Gamma^2/4} \cdot \frac{4 \kappa_\mathrm{ext}}{\kappa_\mathrm{tot}^2 + 4(\omega_p - \omega_r)^2} \frac{P_\mathrm{in}}{\hbar\omega_p}. \label{eq:sup_pump_freq}
\end{align}
During maser operation, the pump frequency $\omega_p$ is tuned to match the resonance frequency of the P1$_0$ spins, i.e., $\omega_p = \omega_s = \omega_0$, but it is off-resonance with the resonator, thus $\omega_p \neq \omega_r$. 
Figure~\ref{figS:rate_eqs}b displays the population difference in the P1+ transition, $\Delta N_\text{P1+}$, plotted against the calculated pump rates $\Gamma_\text{p}$. 
This population difference is extracted using the spin function $K(\omega)$, as discussed in the above section. 
The plot also includes overlayed simulation results from the rate equations Eqs.~\eqref{eq:sup_rateeqs} and \eqref{eq:sup_photon_eq}. 

\subsubsection*{Self-oscillation threshold}
If the photons produced by stimulated emission exceed the resonator losses, the system enters the self-oscillating or free-running maser regime. 
In this regime, the maser is capable of emitting microwave photons into the resonator without the need for an externally applied probe tone. 
The threshold population difference $\Delta N_\mathrm{thr}$ can be calculated from the photon number rate equation given in Eq.~\eqref{eq:sup_photon_eq}~\cite{carmichael_statistical_2013}. 
At the threshold, the two terms in the equation are balanced, hence,
\begin{align}
    0 = - \kappa_\mathrm{tot} n_\mathrm{ss} - \Gamma_s n_\mathrm{ss} \Delta N_\mathrm{thr} + \Gamma_s N_4^\mathrm{thr},
\end{align}
where $N_4^\mathrm{thr}$ is the number of spins in the state 4 at the threshold and $n_\mathrm{ss}$ is the photon number in the steady state in the free-running maser regime. 
Rearranging this equation gives 
\begin{align}
    \Delta N_\mathrm{thr} = -\frac{\kappa_\mathrm{tot}}{\Gamma_s} + \frac{N_4^\mathrm{thr}}{n_\mathrm{ss}}. 
\end{align}
The second term can often be neglected since typically $n_\mathrm{ss} \gg 1$.
Therefore,
\begin{align}
    |\Delta N_\mathrm{thr}| \approx \frac{\kappa_\mathrm{tot}}{\Gamma_s} = \frac{\kappa_\mathrm{tot} \Gamma}{4g_0^2}.
\end{align}
In the case of $Q_\mathrm{ext} \approx 250$, $|\Delta N_\mathrm{thr}|$ is approximately $4.81 \times 10^{14}$, while for $Q_\mathrm{ext} \approx 730$, it is approximately $2.42 \times 10^{14}$. 
These thresholds are shown as the red dot-dashed lines in Fig.~\ref{fig:Fig3}d in the main text and Fig.~S3a and c in the Supplementary information. 


\clearpage

\begin{table}[h]
  \centering
  \caption{Line attenuations}
  \label{tab:line_atts}
  \begin{tabular*}{\textwidth}{@{\extracolsep\fill}lc}
    \toprule
    {Line} & {Attenuation value (dB)} \\
    \midrule
    {Probe} & {53.3} \\
    {Pump, Cancellation} & {27.6} \\
    {Channel 1} & {1.26} \\
    {Channel 2} & {1.50} \\
    \bottomrule
  \end{tabular*}
\end{table}

\begin{table}[h]
  \centering
  \caption{Relaxation rates used in the rate-equation simulation. The values of $\Gamma_{41}$ and $\Gamma_{63}$ are determined from $T_1$ measurements of the P1$_{+}$ and P1$_{-}$ transitions shown in Fig.~\ref{fig:Fig2}f.}
  \label{tab:relax_rates}
  \begin{tabular*}{\textwidth}{@{\extracolsep\fill}lcccc}
    \toprule
    Parameter & {$\Gamma_{41} / 2\pi$ ($\mathrm{s}^{-1}$)} & {$\Gamma_{52} / 2\pi$ ($\mathrm{s}^{-1}$)} & {$\Gamma_{63} / 2\pi$ ($\mathrm{s}^{-1}$)} & {$\Gamma_\mathrm{cr} / 2\pi$ ($\mathrm{s}^{-1}$)} \\
    \midrule
    Value     & $9.6\times10^{-5}$ & $9.6\times10^{-5}$ & $1.5\times10^{-4}$ & $6.0\times10^{-2}$ \\
    \bottomrule
  \end{tabular*}
\end{table}

\begin{table}[h]
  \centering
  \caption{Resonator and spin parameters used in the rate-equation simulations.}
  \label{tab:rate_eqs_params}
  \begin{tabular*}{\textwidth}{@{\extracolsep\fill}ccccccc}
    \toprule
    $Q_\mathrm{ext}$ & $Q_\mathrm{int}$ & $Q_\mathrm{tot}$ & {$\omega_r/2\pi$ (Hz)} & {$\kappa_\mathrm{ext}/2\pi$ (Hz)} & {$\kappa_\mathrm{int}/2\pi$ (Hz)} & {$\Gamma/2 / 2\pi$ (Hz)} \\
    \midrule
    $250$ & $2600$ & $230$ &  {$6.59\times10^{9}$} & $25\times10^6$ & $2.6\times10^6$ & $0.75\times10^6$  \\
    $730$ & $3000$ & $600$ & {$6.86\times10^{9}$} & $9.4\times10^6$ & $2.2\times10^6$ & $0.90\times10^6$ \\
    \bottomrule
  \end{tabular*}
\end{table}



\clearpage

\renewcommand{\figurename}{Fig.}
\renewcommand{\tablename}{Table}
\renewcommand{\thefigure}{S\arabic{figure}}
\renewcommand{\thetable}{S\arabic{table}}
\renewcommand{\theequation}{S\arabic{equation}}
\renewcommand{\thesection}{\arabic{section}}
\renewcommand{\thesubsection}{\thesection.\arabic{subsection}}
\renewcommand{\thesubsubsection}{\thesubsection.\arabic{subsubsection}}
\setcounter{figure}{0}
\setcounter{table}{0}
\setcounter{equation}{0}
\setcounter{section}{0}

\begin{center}
{\large\bfseries Supplementary Information}
\end{center}

\vspace{1em}

\section{Supplementary Methods}
\subsection{Cryogenic and room temperature setups}

In this section, we describe the wiring and setup both inside the dilution refrigerator and at room temperature. 

\subsubsection{Wiring configuration in the dilution refrigerator}
\begin{figure}
    \centering
    \includegraphics[width=0.77\textwidth]{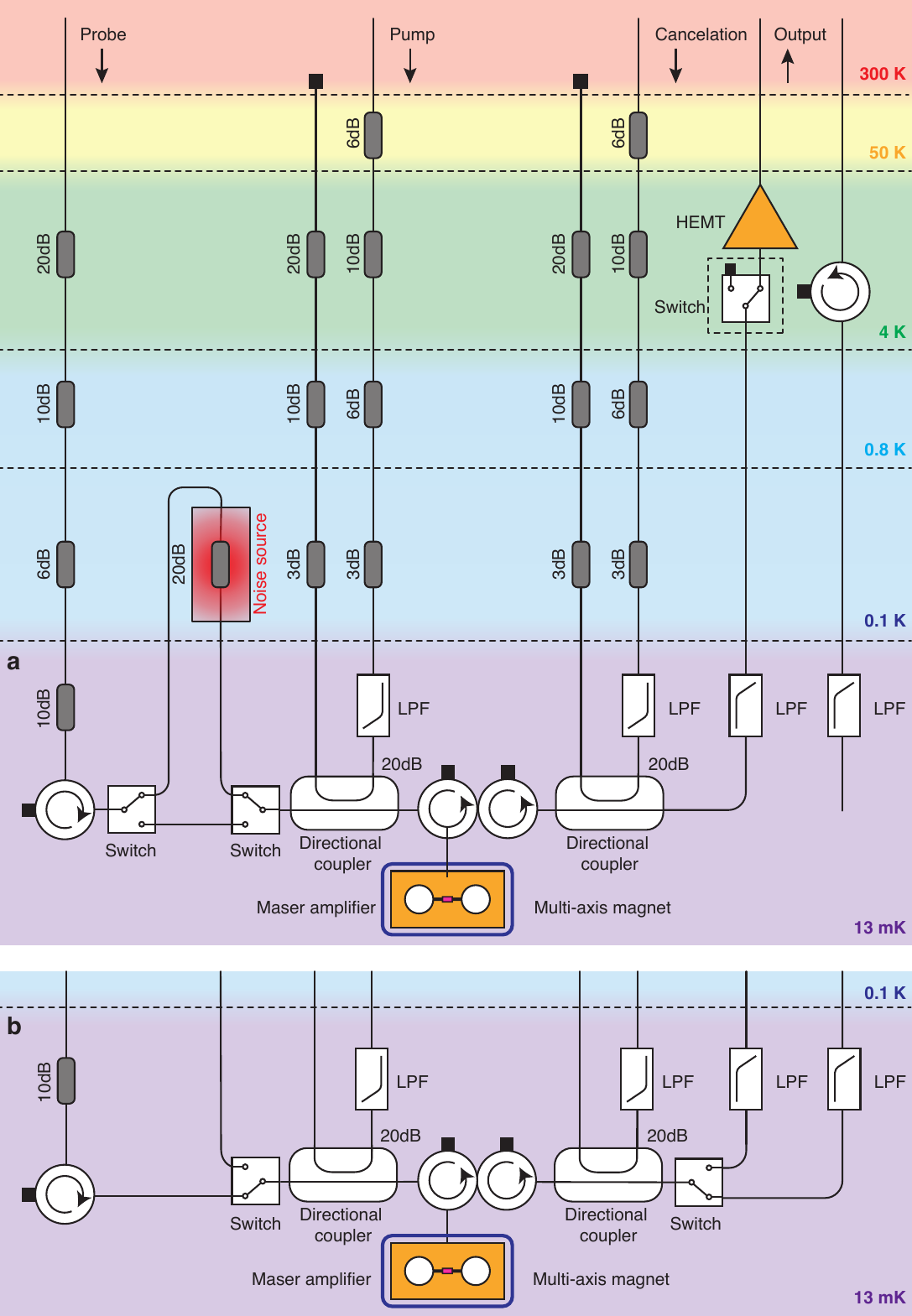}
    \caption{\textbf{Cryogenic wiring and configurations. }
    ``LPF'' and ``IRF'' denote a $7.5\,\mathrm{GHz}$ low pass filter and an infrarad filter, respectively.
    The microwave switch located below the high-mobility transistor amplifier (HEMT) at $4\,\mathrm{K}$ is used only for pulse-ESR (electron spin resonance) measurements. 
    The decibel numbers indicate the values of attenuators (gray round rectangular) and directional couplers. 
    \textbf{a}, Setup at $13\,\mathrm{mK}$ for maser gain and noise temperature measurements. 
    \textbf{b}, Setup at $13\,\mathrm{mK}$ for the saturation measurements. The right-side switch at $13\,\mathrm{mK}$ is used to bypass the HEMT during these measurements. 
    }
    \label{figS:Cryo_setup1}
\end{figure}

The detailed wiring and configuration inside the dilution refrigerator (Bluefors, LD-400) are illustrated in Fig.~\ref{figS:Cryo_setup1}. 
Cryogenic attenuators (XMA, 2082-624X-CRYO series), represented as gray round rectangles, are inserted in each input line at every temperature stage to attenuate thermal noise from the microwave tone injected from room temperature. 
The heavily attenuated ``Probe'' line and moderately attenuated ``Pump'' line are combined at a directional coupler (Marki Microwave, C20-0226) on the mixing chamber stage. 
The labels ``LPF'' and ``IRF'' represent low-pass filters (Marki Microwave, FLP-0750) and homemade infrared filters using iron-loaded epoxy Eccosorb CR110, respectively. 
A HEMT (High Electron Mobility Transistor) amplifier (Low Noise Factory, LNA-LNC4\_8C) was installed on the $4\,\text{K}$ plate. 

The calibrated noise source~\cite{simbierowicz_characterizing_2021} is used for the noise temperature measurements. 
While the noise temperature is heated maximally up to $\approx 4.5\,\text{K}$, the base temperature of the mixing chamber plate slightly increases up to $\approx 19\,\text{mK}$. 

Two microwave switches (Radiall, R573423600) were used to check the total attenuation of the coaxial cables to and from the noise source, which turned out to be negligible. 
The microwave switch (Analog Devices Inc., HMC547ALP3E), depicted inside the dashed box in Fig.~\ref{figS:Cryo_setup1}, was installed and used only for pulse electron spin resonance (pulse-ESR) measurements to protect the HEMT. 
Further details are described in Sec .~\ref {sec:sup_edfs}.

\subsubsection{Pump cancellation}
\begin{figure}
    \centering
    \includegraphics[width=1\textwidth]{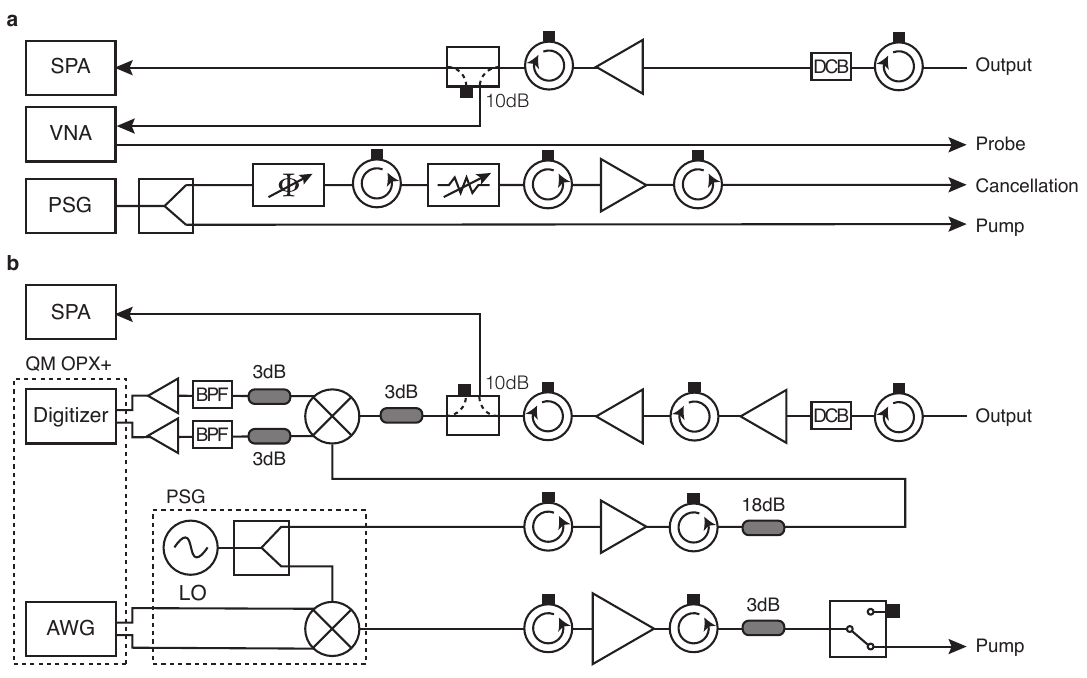}
    \caption{\textbf{Room temperature measurement setup.} ``BPF'' and ``DCB'' denote a $94 - 108\, \mathrm{MHz}$ band pass filter and a DC block, respectively. \textbf{a}, Setup for masing measurement. \textbf{b}, Setup for pulse ESR measurement.}
    \label{figS:RT_setup}
\end{figure}

The probe tone was fed into the fridge through the line labeled ``Probe'', while the pump tone was sent through the line ``Pump''. 
To minimize the impact of the pump tone on the HEMT and the following receiver circuits, a cancellation microwave tone was introduced through a dedicated microwave input line labeled ``Cancellation'' (see also Fig.~\ref{figS:RT_setup}). 
When operating the maser amplifier, a microwave tone was generated by a signal generator (PSG; Keysight Technologies, E8267D) for both pump and cancellation. 
This tone was then split by a power splitter (Mini-Circuits, ZC2PD-01263-S+), as shown in Fig.~\ref{figS:RT_setup}a. 
On one branch, the cancellation tone was adjusted using a voltage-controlled phase shifter (Qorvo, CMD297P34) and a voltage-controlled variable attenuator (Pulsar Microwave, AAT-22-479A/7S). 
The amplitude and phase of this cancellation tone were adjusted to destructively interfere with the reflected pump tone through a directional coupler placed after the two circulators (Quinstar, QCY-G040082AS), resulting in suppression of the pump power by at least $50\,\text{dB}$.

\subsubsection{Setup for maser gain and noise temperature measurement}

The reflection coefficient $r(\omega)$ of the resonator was measured using a vector network analyser (VNA; Keysight, N5230C). 
The data presented in Figs.~2f and 3a, b, and d in the main text were measured at a power of $-110\,\mathrm{dBm}$, corresponding to a mean intra-resonator photon number of approximately $10^{6}$, which is sufficiently lower than the number of inverted spins, $\Delta N_{P1+} \approx 4\times10^{14}$, ensuring minimal impact on the spin saturation. 
Although this power level almost reaches the typical saturation power of traveling-wave parametric amplifiers~\cite{esposito_perspective_2021}, it remains well below the saturation power of our maser amplifier as discussed in Fig.~4c in the main text. 

Noise power spectra were measured by a spectrum analyser (SPA; Keysight Technologies; N9010A). 
The output and input ports of the VNA are connected to the Probe and Output ports of the fridge via directional couplers (input: Pulsar Microwave, CS20-09-436/9; output: Marki, C10-0226). 
A signal generator (SG, Keysight Technologies, E8247C) generates a microwave probe tone during the signal-to-noise ratio (SNR) measurements, as presented in Fig.~4a. 

\subsubsection{Setup for compression measurement}
Figure~\ref{figS:Cryo_setup1}b displays the wiring configuration on the mixing chamber plate used to measure the $1\,\text{dB}$ compression points, as presented in Fig.~4C in the main text.  
To ensure the HEMT and subsequent components were not saturated, we bypassed the HEMT using a microwave switch. 

\subsubsection{Setup for pulse electron spin resonance measurements}
In the pulse ESR measurement detailed in Sec.~\ref{sec:sup_edfs}, we use an FPGA-based integrated measurement system (Quantum Machines, OPX+) to generate pulse envelopes, control microwave switches, and detect spin echo signals. 
Gaussian-shaped pulse envelope signals, modulated at $100\,\mathrm{MHz}$, are generated by the OPX arbitrary waveform generator (AWG). 
We use the single sideband modulation technique, where a baseband microwave tone (``LO'' in Fig.~\ref{figS:RT_setup}b) is mixed with the pulse envelopes with an internal IQ mixer of the PSG. 
These pulses are subsequently amplified by a high-power amplifier (Cree, CMPA601C025F) and fed to the resonator through the ``Pump'' line. 
To minimize microwave leakage noise from the amplifier, a semiconductor-based microwave switch (Analog Devices Inc., ADRF5019-EVALZ) is installed immediately after the amplifier's output. 
Similarly, another semiconductor-based microwave switch (Analog Devices Inc., HMC547ALP3E) is installed to protect the HEMT amplifier from the pulses. 

For echo signal detection, the signals from the output port of the refrigerator are amplified by a series of room-temperature amplifiers (B\&Z Technologies, BZ-02000800-090826-182020, and Mini-Circuits, ZVA-183+) and then down-converted using an IQ mixer (Marki, IQ-0307L). 
The local oscillator (LO) tones are extracted from a portion of the PSG’s locking signal (option HCC). 
After demodulation, the signals are passed through band-pass filters (Mini-Circuits, SBP-101+), and then amplified by a preamplifier (SRS, SR445A). 
Finally, they are detected by the OPX's digitizer, which processes the signals digitally for heterodyne detection.

\section{Supplementary Discussion}
\subsection{Self-oscillation regime}
\begin{figure}
    \centering
    \includegraphics[width=0.8\linewidth]{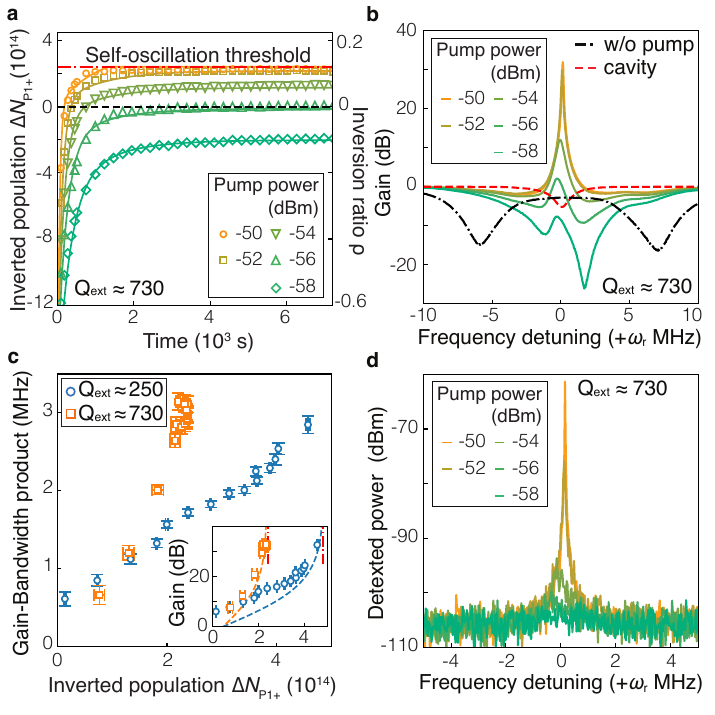}
    \caption{
    \textbf{Maser amplifier gain and bandwidth with $Q_\mathrm{ext}\approx730$}.
    \textbf{a}, Inverted spin population $\Delta N_\mathrm{P1_+}$ between the lower and upper P1$_{+}$ states over time under various pump powers for an external quality factor of $Q_\text{ext} \approx 730$. 
    Open symbols represent the experimental data, while solid curves represent simulations. 
    \textbf{b}, Maser gain spectra under various pump powers. 
    The horizontal axis represents the detuning from the central resonator frequency $\omega_r/2\pi = 6.595\,\text{GHz}$. 
    Black and red dashed lines represent the spectrum measured without the pump and with the spins off resonance, respectively.
    \textbf{c}, Gain–bandwidth product versus inverted spin population. 
    The inset shows the gain versus inverted spin population, where the open symbols and dashed curves are the experimental data and theoretical curves, respectively (Methods in the main text).
    \textbf{d}, Power spectrum of the maser amplifier with $Q_\mathrm{ext}\approx730$. 
    At $P_\mathrm{pump} = -50\,\mathrm{dBm}$, the system enters the self-oscillation regime.
    }
    \label{fig:Qe730}
\end{figure}

We measured the gain spectrum of the maser amplifier with an external quality factor of $Q_\mathrm{ext}\approx 730$.
Figure~\ref{fig:Qe730}a shows the same type of measurement as in Fig.~3a in the main text ($Q_\mathrm{ext}\approx 250$), where the system reaches the self-oscillation threshold.
This is evidenced by the saturation of the population at $\Delta N_\mathrm{P1_+}\approx2.4\times10^{14}$, which aligns well with the calculated threshold indicated by the red dot-dashed horizontal line.
The maser gain spectrum measured using a vector network analyzer is presented in Fig.~~\ref{fig:Qe730}b. 
The gain-bandwidth product and gain as a function of the inverted population $\Delta N_\mathrm{P1_+}\approx2.4\times10^{14}$ are also shown in Fig.~\ref{fig:Qe730}c and its inset.
Figure~~\ref{fig:Qe730}d displays the power spectrum of the maser amplifier under various pump powers.

\subsection{Echo-detected field-sweep spectroscopy}
\label{sec:sup_edfs}
To characterise the impurity spins in our diamond, specifically the ``fast-relaxing spins,'' we conducted an echo-detected field-sweep (EDFS) measurement using the setup depicted in Fig.~\ref{figS:RT_setup}b, with the pulse sequence illustrated in Fig.~\ref{figS:pulseESR}a. 
A practical challenge arises from the ``long relaxing spins'', namely the P1 and NV$^{-}$ centres, which exhibit very long relaxation times ranging from approximately $\sim 10^3$ to $\sim 10^4$ seconds. 
Additionally, the substantial quantity of these impurity spins may hinder the signals from the ``fast-relaxing spins'', which turned out to be about two orders of magnitude less abundant. 
To mitigate this, we first saturated the P1 and NV$^{-}$ centres using a 200-ms-long saturation pulse with frequency modulation of $\pm 10\, \mathrm{MHz}$ centred at the resonator frequency of $\mathrm{6.788\, GHz}$ in this cooldown. 
This allowed us to primarily detect the spin echo signals from the ``fast-relaxing spins''. 
This approach also enabled optimal adjustment of the room temperature amplifier chain to utilize the full dynamic range of the digitizer. 
After a waiting time of $T_W = 100\,\mathrm{ms}$, the echo signal was detected using a Hahn-echo sequence consisting of $\pi/2$ and $\pi$ pulses. 

The results of the EDFS measurement are shown in Fig.~\ref{figS:pulseESR}b. 
Despite the saturation pulse, several peaks associated with the P1 and NV$^{-}$ centres remain visible. 
This is attributed to imperfections in the saturation pulse and partial population recovery during the $100\,\mathrm{ms}$ wait time. 
Nevertheless, a distinct ``broad shoulder'' feature is observed, extending from around the P1$_{-}$ peak at approximately $245\,\text{mT}$ to around $260\,\text{mT}$. 
We fitted the data with a sum of nine Lorentzian functions. 
These components are also indicated by short arrows at the top of Fig.~\ref{figS:pulseESR}b.
The broad signal stemming from the ``fast-relaxing spins'' is represented by a solid red curve. 

\subsection{Characterization of ``fast-relaxing spins''}
\label{sec:sup_T1meas}

\begin{figure}
    \centering
    \includegraphics[width=0.8\textwidth]{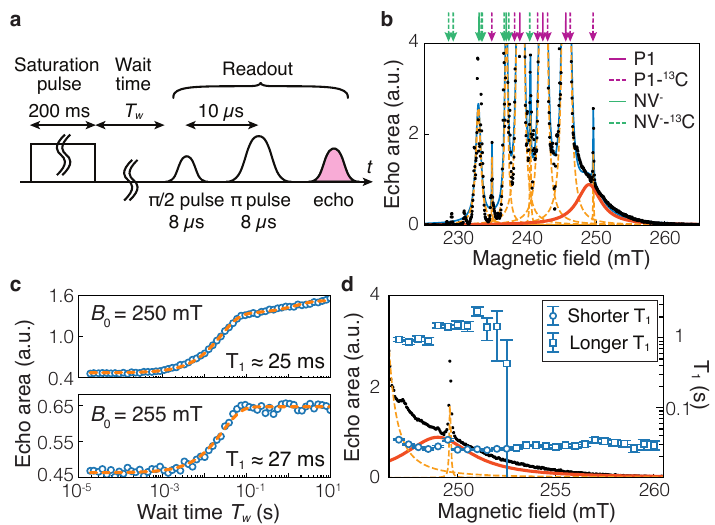}
    \caption{\textbf{Electron spin resonance spectroscopy and relaxation time measurements.}
    \textbf{a}, Pulse sequence used for the echo-detected field-sweep (EDFS) spectroscopy and saturation recovery measurements. 
    The wait time $T_\text{W}$ was fixed at 100 ms for EDFS spectroscopy and was varied for $T_1$ measurements. 
    \textbf{b}, EDFS signals as a function of the magnetic field. 
    Black points depict experimental data, while the blue solid curve indicates the fit using nine Lorentzian functions. 
    Orange dashed curves correspond to fitting results for known defects in diamond, i.e., P1 and NV$^-$ centres, and their hyperfine transitions with $^{13}$C centres at the nearest neighbors. 
    The red solid curve represents fast-relaxing spins. 
    Calculated resonant magnetic field values for known defects are indicated by short arrows at the top.
    \textbf{c}, Results of saturation recovery measurements. 
    Echo area as a function of wait time $T_W$ at a magnetic field of $B_0 = 250\, \mathrm{mT}$ (upper panel) and $B_0 = 255\, \mathrm{mT}$ (lower panel) are plotted. 
    Orange dashed lines show the result of bi-exponential fitting in the upper panel and single-exponential fitting in the lower panel.
    \textbf{d}, $T_1$ as a function of the magnetic field, with the EDFS spectroscopy data overlaid. Circular and square markers represent the shorter and longer $T_1$ components obtained from bi-exponential fits. 
    Data obtained above $253\,\text{mT}$ can be fitted with a single exponential. 
    }
    \label{figS:pulseESR}
\end{figure}

We measured the longitudinal relaxation times $T_1$ of the ``fast-relaxing spins'' under constant magnetic fields ranging from $247\, \mathrm{mT}$ to $260 \, \mathrm{mT}$ using the saturation recovery pulse sequence~\cite{schweiger_principles_2001}. 
The results are shown in Fig.~\ref{figS:pulseESR}c and d. 
Near the transition of P1$_{-}$, the relaxation curves exhibit bi-exponential decays, as presented in the upper panel of Fig.~\ref{figS:pulseESR}c.  
In contrast, in the fields higher than $253 \, \mathrm{mT}$, the relaxation curves can be fitted by a single exponential decay (bottom panel of Fig.~\ref{figS:pulseESR}c). 
This is attributed to the cross-relaxation effects with the P1$_{-}$ transition and their $^{13}$C hyperfine-coupled transition. 

\begin{figure}
    \centering
    \includegraphics[width=0.8\textwidth]{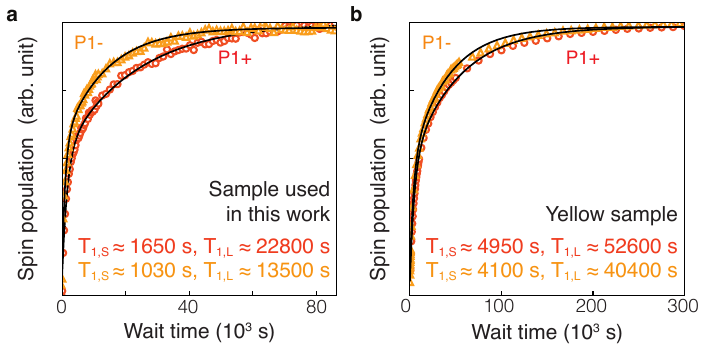}
    \caption{\textbf{Spin relaxation measurements of two different diamond samples.} 
    \textbf{a}, Results of the sample used in this work. 
    The data were fitted using a bi-exponential function, and the two relaxation times, i.e., the shorter $T_\text{1,S}$ and longer $T_\text{1,L}$, are displayed. 
    \textbf{b}, Results of a yellow diamond sample. 
    }
    \label{figS:P1_relaxation}
\end{figure}

As noted in the main text, we have not been able to rigorously identify the specific spin species constituting the ``fast-relaxing spins.'' 
Nevertheless, the evidence presented below suggests that these may be neutrally-charged nitrogen-vacancy (NV$^0$) centres. 
First, we measured $T_1$ of P1 centres in a type Ib yellow diamond, an as-grown HPHT sample purchased from Sumitomo Electric Industries, Ltd., containing approximately $50$ to $100\,\mathrm{ppm}$ of P1 centres, at $10\,\mathrm{mK}$. 
The results, displayed in Fig.~\ref{figS:P1_relaxation}b, show spin relaxation times exceeding $4 \times 10^3$ seconds in the short $T_1$, about four or five times longer than those measured in the sample used for the maser experiments as presented in Fig.~\ref{figS:P1_relaxation}a. 
Similarly, the longer $T_1$ also shows $5.2 \times 10^4$ and $4 \times 10^4$ seconds. 
These imply that the electron irradiation, i.e., the introduction of vacancies into our diamond crystal, likely created the ``fast-relaxing spins'' that exhibit the broad spin distribution (Fig.~\ref{figS:pulseESR}b), subsequently impacting the transitions of P1 centres. 

Vacancy clusters~\cite{iakoubovskii_dominant_2002} initially appeared to be possible candidates. 
However, these defects typically exhibit spin transitions with much narrower linewidths (significantly less than $1\, \text{mT}$) at temperatures below $100\,\text{K}$~\cite{iakoubovskii_dominant_2002}, which stands in stark contrast to the broad transition observed for the fast-relaxing spins in our diamond crystal. 
On the other hand, similar to negatively-charged silicon-vacancy (SiV$^{-}$) centres~\cite{meesala_strain_2018}, the spins of NV$^0$ centres are sensitive to crystal strain~\cite{baier_orbital_2020, kurokawa_coherent_2024}. 
Such strain becomes significant in diamond crystals containing defect concentrations exceeding $\approx 1\,\mathrm{ppm}$~\cite{liu_spectral_2021, acosta_electromagnetically_2013}, which is indeed the case for the diamond crystal used in our experiments. 

\subsection{SNR measurement with varying input noise}
To further confirm that the SNR change observed in Fig.~4a is associated with the low-noise maser amplifier itself, and not with artefacts such as cavity or spin absorption, we measured the signal-to-noise ratio (SNR) as a function of the input noise $n_\mathrm{in}$. 
As shown in Fig.~\ref{figS:SNR_improvement}, the SNR change decreases monotonically with increasing $n_\mathrm{in}$. 
The expected SNR change was calculated using Eqs.~(7) and (8) in Methods. 
The resulting predictions for different assumed maser added-noise values are shown as the blue band ($n_\mathrm{maser}+0.5 = 3$, corresponding to the measured value) and the orange band ($n_\mathrm{maser}+0.5 = 9.3$, corresponding to the HEMT noise) in Fig.~\ref{figS:SNR_improvement}.
The measured data fall within the blue band, providing a strong internal consistency check that the maser added noise is significantly smaller than the noise contribution of the subsequent HEMT amplifier. 

\subsubsection{Analytical model for SNR change}\label{sec:sup_snr_improvement}
In Fig.~4a, the noise powers at the noise floor are written as
\begin{align}
    P_\mathrm{on}^\mathrm{nf} = P_\mathrm{on}, \quad
    P_\mathrm{off}^\mathrm{nf} = P_\mathrm{off},
\end{align}
where $P_\mathrm{on}$ and $P_\mathrm{off}$ correspond to Eqs.~(7) and (8) in the ``Method'', respectively.

Similarly, the noise powers at the peak are expressed as
\begin{align}
    P_\mathrm{on}^\mathrm{pk} &= P_\mathrm{on} + G_\mathrm{maser}G_\mathrm{sys} n_\mathrm{probe} \cdot \hbar\omega \delta f_\text{BW}, \\
    P_\mathrm{off}^\mathrm{pk} &= P_\mathrm{off} + G_\mathrm{sys} n_\mathrm{probe} \cdot \hbar\omega \delta f_\text{BW},
\end{align}
where $G_\mathrm{sys} = G_\mathrm{HEMT}\beta_1\beta_2$, and $n_\mathrm{probe}$ denotes the input photon number of the probe tone.

The SNR change $\mathcal{I}$ can be written as
\begin{align}
    \mathcal{I} = \frac{\mathrm{SNR}_\mathrm{on}}{\mathrm{SNR}_\mathrm{off}} = \frac{P_\mathrm{on}^\mathrm{pk}}{P_\mathrm{on}^\mathrm{nf}} \cdot \frac{P_\mathrm{off}^\mathrm{nf}}{P_\mathrm{off}^\mathrm{pk}}.
\end{align}
Using the above equations, we obtain
\begin{align}\label{eq:chap6_snr_improvement}
    \mathcal{I} &= \frac{P_\mathrm{on} + G_\mathrm{maser}G_\mathrm{sys} n_\mathrm{probe}}{P_\mathrm{on}} \cdot \frac{P_\mathrm{off}}{P_\mathrm{off} + G_\mathrm{sys} n_\mathrm{probe}} = \frac{1 + \mathcal{A}}{1+\mathcal{B}}
\end{align}
where
\begin{align}
    \mathcal{A} = \frac{G_\mathrm{maser}G_\mathrm{sys} n_\mathrm{probe}}{P_\mathrm{on}}, \quad
    \mathcal{B} = \frac{G_\mathrm{sys} n_\mathrm{probe}}{P_\mathrm{off}}.
\end{align}

\begin{figure}
    \centering
    \includegraphics[width=0.8\linewidth]{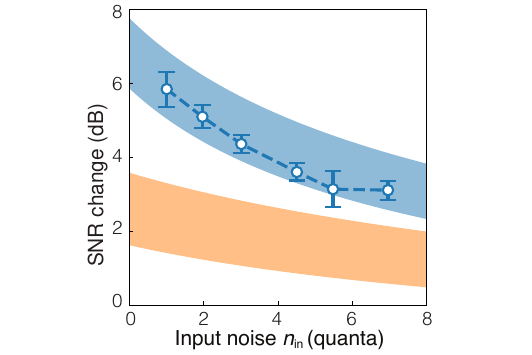}
    \caption{SNR change as a function of input noise $n_\mathrm{in}$.
             Blue and orange bands indicate the predicted improvement values for $n_\mathrm{maser}+0.5=3$ (measured value) and $n_\mathrm{maser}+0.5=9.3$ (corresponding to the HEMT noise), respectively, with an uncertainty of $\beta_2=-3.5$ to $-1.5\,\mathrm{dB}$ (see Supplementary Discussion Sec.~2.4.1)}
    \label{figS:SNR_improvement}
\end{figure}

Although the probe tone is weak, its photon number $n_\mathrm{probe}$ is still much bigger than that of the input noise $n_\mathrm{in}$. ($P_\mathrm{probe}\approx-140\,\mathrm{dBm}$, $n_\mathrm{probe} \approx 10^{4}$, $n_\mathrm{in} \approx 1-10$).
Consequently, $\mathcal{A}$ and $\mathcal{B}$ satisfy  $\mathcal{A}, \mathcal{B} \gg 1$.
Therefore, Eq.~\eqref{eq:chap6_snr_improvement} can be approximated as
\begin{align}
    \mathcal{I} &\approx \frac{\mathcal{A}}{\mathcal{B}} = \frac{G_\mathrm{maser}P_\mathrm{off}}{P_\mathrm{on}} \label{eq:chap6_snr_impr_approx1}\\
    &\approx \frac{G_\mathrm{maser} \left[n_\mathrm{in,eff} + \frac{1}{\beta_2}\left(n_\mathrm{HEMT} + \frac{1}{2}\right)\right]}
    {G_\mathrm{maser}n_\mathrm{in,eff} + (G_\mathrm{maser}-1)\left(n_\mathrm{maser} + \frac{1}{2}\right)+ \frac{1}{\beta_2}\left(n_\mathrm{HEMT} + \frac{1}{2}\right)}.\label{eq:chap6_snr_impr_approx2}
\end{align}
where $n_{\mathrm{in, eff}} = \beta_1 \left(n_\mathrm{in} + 1/2\right) + (1-\beta_1) \left(n_0 + 1/2\right)$ is the effective input noise.
Here, we approximate $G_\mathrm{HEMT}-1 \approx G_\mathrm{HEMT}$ because the HEMT operates in the high-gain regime ($G_\mathrm{HEMT}\approx10^{4}$). 
We used $G_\mathrm{maser} = 20\,\mathrm{dB}$ and $n_\mathrm{HEMT} = 9.3$ quanta, and assumed $\beta_2$ to be between $-3.5$ and $-1.5\,\mathrm{dB}$ to calculate the expected SNR change shown in Fig.~\ref{figS:SNR_improvement}.

Although Eq.~\eqref{eq:chap6_snr_impr_approx1} appears to contain no explicit information about the peak values, dividing $P_\mathrm{on}$ by $G_\mathrm{maser}$ corresponds, in the dB scale, to shifting the noise floor in the ``pump on'' condition by the same amount as the peak-height difference.
In other words, this operation is equivalent to the downward vertical shift applied to the ``pump on'' data in Fig.~4a in the main text, and captures the resulting SNR change.


\end{document}